\documentclass[10pt,dvips,a4paper]{article}
\include{commands}
\include{journals}
\usepackage{amssymb}
\usepackage{amsmath}
\usepackage{setspace}
\usepackage[numbers]{natbib}
\usepackage[pdftex]{graphicx,xcolor}

\renewcommand{\title}{Three-Dimensional Network Model for Coupling~of~Fracture and Mass Transport in Quasi-Brittle Geomaterials}

\begin{document}

\begin{center} \begin{LARGE} \textbf{\title} \end{LARGE} \end{center}

\begin{center} Peter Grassl$^{1*}$ and John Bolander$^{2}$\\
  $^{1}$~School of Engineering, University of Glasgow, Glasgow, UK\\
  $^{2}$~Department of Civil and Environmental Engineering, University of California, Davis, U.S.A.

  $^*$Corresponding author: Email: peter.grassl@glasgow.ac.uk, Phone: +44 141 330 5208
  
\end{center}

\section*{Abstract}
Dual three-dimensional networks of structural and transport elements were combined to model the effect of fracture on mass transport in quasi-brittle geomaterials. Element connectivity of the structural network, representing elasticity and fracture, was defined by the Delaunay tessellation of a random set of points. The connectivity of transport elements within the transport network was defined by the Voronoi tessellation of the same set of points. A new discretisation strategy for domain boundaries was developed to apply boundary conditions for the coupled analyses. The properties of transport elements were chosen to evolve with the crack opening values of neighbouring structural elements. Through benchmark comparisons involving non-stationary transport and fracture, the proposed dual network approach was shown to be objective with respect to element size and~orientation.

\section{Introduction}
The influence of fracture and mass transport affects physical processes that govern many engineering applications, such as deterioration of construction materials and performance of waste barriers.
In these applications, it is important to consider the influence of fracture induced pathways for the ingress of fluids.
Modelling the discrete crack formations, and the mass transport along these cracks and through the surrounding uncracked material, is challenging.
Models for this coupling are commonly based on continuum mechanics combined with a discrete representation of cracks \citep{RoeMooDeProCar06,SegCar08,SegCar08a,CarGra12,YaoLiKee15}.
Alternatively, discrete approaches, such as discrete element method, lattice and network models, have been proposed to model these processes \citep{SadMie97,ChaPicKhePij05,NakSriYas06,WanSodUed08,Gra09,SavPacSch13,AsaHouBir14,GraFahGal15,MarDerMoh15,DamDetCun15}.

\begin{figure}[ht]
\centering
      \includegraphics[width=14cm]{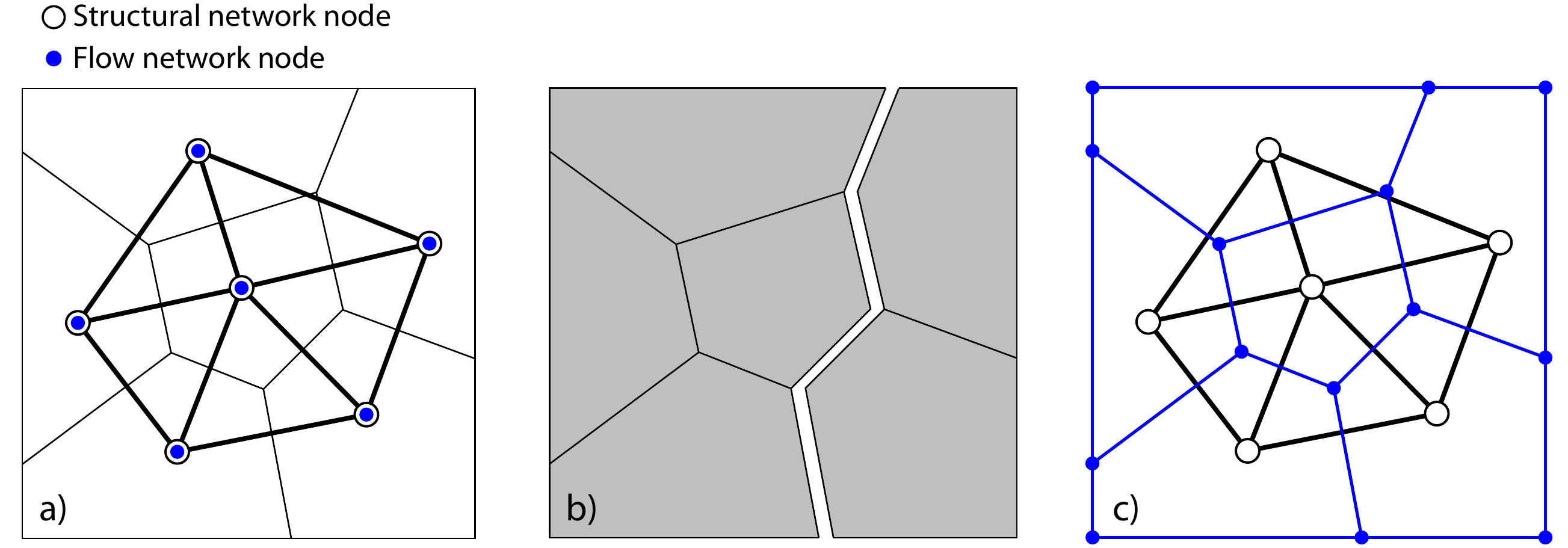}
    \caption{Network models for coupled problems: (\textbf{a}) common approach in which the structural and transport network nodes are coincident. Both structural and transport elements are on the Delaunay edges; (\textbf{b}) simulated crack in structural network; and (\textbf{c}) improved approach in which transport elements are on the Voronoi edges and therefore aligned with potential cracks.}
\label{fig:2D}
\end{figure}

One network approach, based on the Delaunay tessellation of a random set of points, has been shown to be suitable for modelling fracture \citep{BolSai98,BolSuk05,GraJir10,GraDav11} and mass transport \citep{BolBer04}, providing mesh insensitive results. In this approach, the physical processes are modelled by a multi-dimensional network of one-dimensional elements, which are placed on the Delaunay edges (Figure~\ref{fig:2D}a); the element properties are determined by the corresponding Voronoi tessellation.
The nodes of the elements of structural and transport network models coincide, which is suitable for modelling the coupling of continuum fields. However, once cracks are formed, the transport elements in this approach are orientated perpendicular to the crack path, which is aligned with the mid-cross-section of the structural elements (Figure~\ref{fig:2D}b). This misalignment of the transport elements with the crack path complicates the modelling of crack-assisted transport and its dependence on crack opening.
To resolve this deficiency, several researchers~\citep{NakSriYas06,Gra09,Saka12,GraFahGal15} have placed transport elements on the Voronoi edges, whereas the structural elements remain on the Delaunay edges (Figure~\ref{fig:2D}c). With this dual network approach, the influence of fracture on transport is more naturally represented, since the transport elements are aligned with the potential crack directions. So far, most of this work was either limited to 2D or did not provide coupling between fracture and transport.

This work proposes a three-dimensional dual network approach for modelling fracture and mass transport. Structural elements are placed on the edges of Delaunay tetrahedra and transport elements are placed on the edges of Voronoi polyhedra. Special attention is given to the discretisation of the dual networks at domain boundaries. Simple geometric relationships based on Voronoi and Delaunay tessellations are proposed for describing the change of permeability as a function of crack opening. By a series of benchmarks, it has been demonstrated that the present approach can describe fracture, transport, and the increase of permeability due to fracture mesh-independently.
Fracture is modelled by means of a cohesive-frictional approach, which is suitable for geomaterials, such as concrete and rocks in which the size of the fracture process zone is large compared to the size of the structure. Transport is modelled by means of Darcy's flow equation. The proposed model is designed to describe the effect of cohesive fracture on conductivity.

\section{Network Approach}
The new network approach uses one-dimensional elements connected in a three-dimensional network to describe continuum fields as well as evolutions of discontinuities in the form of fracture process zones.
In the present section, the discretisation and mechanical equations of the structural and transport parts are discussed.
At the end of each section, the input parameters for the individual parts are presented.

\subsection{Discretisation} \label{sec:discretisation}
\begin{figure}[ht]
\centering
    \begin{tabular}{ccc}
      \includegraphics[width=5.5cm]{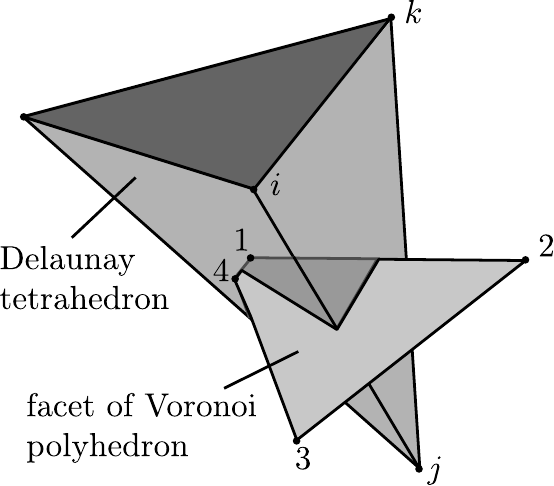} & \includegraphics[width=5cm]{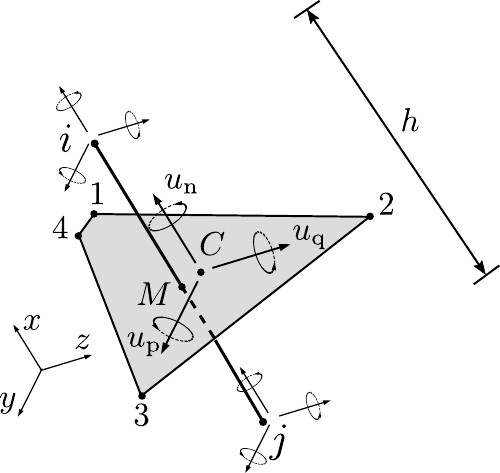} & \includegraphics[width=3cm]{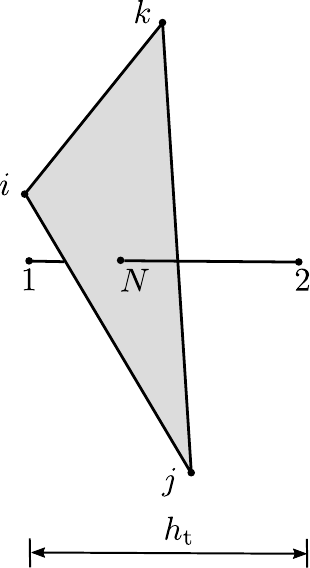}\\
      (\textbf{a}) & (\textbf{b}) & (\textbf{c})
    \end{tabular}
    \caption{Spatial arrangement of structural and transport elements of the 3D transport-structural network approach:~(\textbf{a}) geometrical relationship between Delaunay and Voronoi tessellations; (\textbf{b})~structural element with cross-section defined by the associated Voronoi facet; and (\textbf{c}) transport element with cross-section defined by the associated Delaunay facet.}
\label{fig:3DTess}
\end{figure}

The dual network approach is based on the  Delaunay and Voronoi tessellations of a set of points placed randomly within the domain.
The points are placed sequentially while enforcing a minimum distance $d_{\rm{min}}$ between all points; trial points that fail the minimum distance criterion are rejected.
The Delaunay tessellation decomposes the domain into tetrahedra whose vertices coincide with the randomly placed points; the Voronoi tessellation divides the domain into polyhedra associated with the random points~\citep{OkaBooSug00}.
These geometrical arrangements of Delaunay and Voronoi tessellations are used to define the structural and transport elements.
Figure~\ref{fig:3DTess}a shows a Delaunay tetrahedron and the Voronoi facet associated with Delaunay edge $i$--$j$.
The structural elements are placed on the Delaunay edges with their mid-cross-sections defined by the facets of the Voronoi polyhedra (Figure~\ref{fig:3DTess}b).
Analogous to the structural network, the transport elements are placed on the edges of the Voronoi polyhedra, with their cross-sections formed by the facets of the Delaunay tetrahedra (Figure~\ref{fig:3DTess}c).

\begin{figure}[ht]
\centering
    \begin{tabular}{cc}
      \includegraphics[width=6cm]{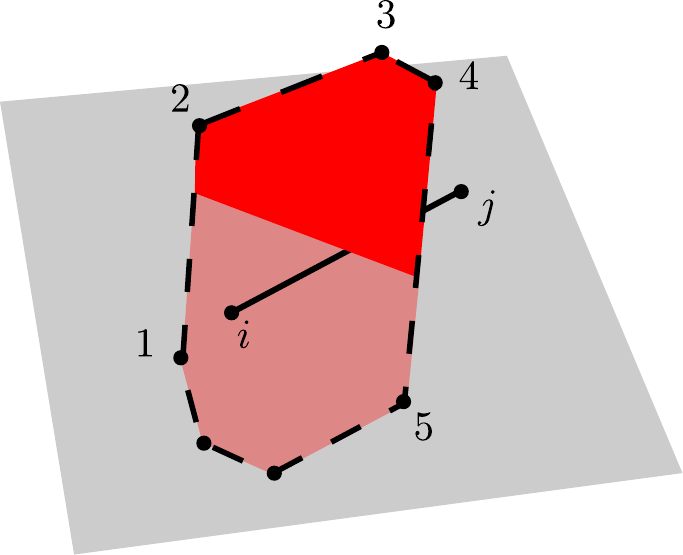} & \includegraphics[width=6cm]{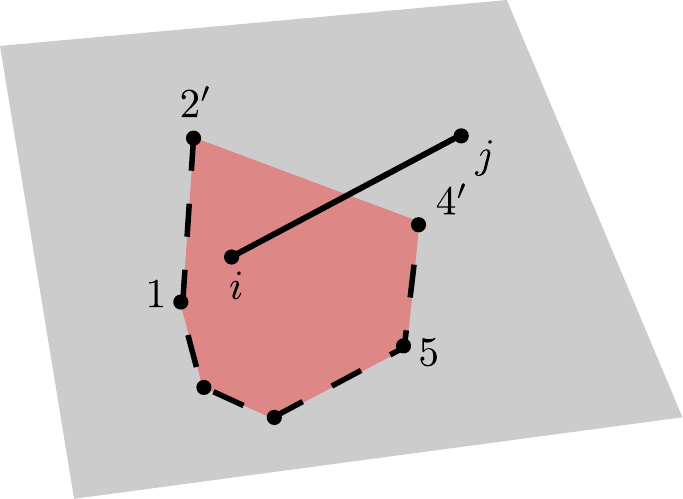}\\
      (\textbf{a}) & (\textbf{b})
    \end{tabular}
  \caption{Discretisation of domain boundaries: (\textbf{a}) Voronoi facet of Delaunay edge $i$--$j$ located on the surface of the domain after initial tessellation; and (\textbf{b}) modified arrangement used for definition of transport nodes and elements.}
\label{fig:tessBound}
\end{figure}

The discretisation of boundaries of the domain requires special attention.
The procedure used in this work is illustrated in Figure~\ref{fig:tessBound}.
Prior to the sequential, random filling of points within the domain, points are placed randomly on the domain surfaces. The minimum distance criterion is enforced during the placement of all of these points. Each interior point is then mirrored with respect to all surfaces of the domain, similar to the procedure of Yip et al.~\cite{YipMohBol05}.
The tessellations for this set of random points results in Delaunay edges located on the domain surfaces, with their corresponding Voronoi facets traversing the domain boundaries as shown in Figure~\ref{fig:tessBound}a.
Here, Delaunay edge $i$--$j$ 
lies on the surface of the domain. Furthermore, Voronoi vertices $1$ and $5$ are inside, and $2$, $3$ and $4$ are outside the domain.
In~constructing the transport network, the Voronoi edges within the domain are retained.
For edges that cross a surface, only the portion within the domain is kept. For example, edges 1--2 and 4--5 become edges 1--2$^\prime$ and 4$^\prime$--5, respectively, where nodes 2$^\prime$ and 4$^\prime$ lie on the surface (Figure~\ref{fig:tessBound}b). These truncated edges define transport elements that are perpendicular to the surface. The modified set of Voronoi edges defines the mid-cross-section of the structural element associated with nodes $i$ and $j$.

Information exchange between the structural and transport networks is based on the geometrical relationship between neighbouring elements. Herein, a one-way coupling is considered, in which crack openings supplied by the structural network affect the conductivity of the associated transport elements. Details regarding this coupling are provided in Section \ref{sec:transElem}. The input parameter for the discretisation is the minimum distance $d_{\rm min}$ which controls the average lengths of structural and transport elements.

\subsection{Structural Network Model}\label{sec:structural}
For the 3D structural analysis, the equilibrium equation for the quasi-static case without body forces \citep{Strang86} is
\begin{equation} \label{eq:stressEqui}
\nabla \boldsymbol{\sigma}^{\rm c} = \boldsymbol{0},
\end{equation}
where $\nabla$ is the divergence operator and $\boldsymbol{\sigma}^{\rm c}$ is the continuum stress.
This equilibrium equation is approximated by a network of structural elements.

\subsubsection{Structural Element}
The discrete version of Equation (\ref{eq:stressEqui}) for the structural element shown in Figure~\ref{fig:3DTess}b is
\begin{equation}
\mathbf{K} \mathbf{u}_{\rm e} = \mathbf{f}_{\rm s},
\end{equation}
where $\mathbf{K}$ is the stiffness matrix, $\mathbf{u}_{\rm e}$ are the vector of degrees of freedom and $\mathbf{f}_{\rm s}$ are the acting forces.~The~formulation of the structural element is presented in the local coordinate system, i.e., the coordinate system ($x$, $y$ and $z$) of the nodal degrees of freedom coincides with the coordinate system \mbox{($n$, $p$ and $q$)} of the quantities used for evaluating the constitutive response.
Each node has three translational \mbox{($u_{\rm x}$, $u_{\rm y}$ and $u_{\rm z}$)} and three rotational ($\phi_{\rm x}$, $\phi_{\rm y}$ and $\phi_{\rm z}$) degrees of freedom.~The degrees of freedom of a structural element with nodes $i$ and $j$ are grouped in translational and rotational parts as \mbox{$\mathbf{u}_{\rm e} = \left\{\mathbf{u}_{\rm t}^T, \mathbf{u}_{\rm r}^T \right\}^T$}, where $\mathbf{u}_{\rm t} = \left\{ u_{\rm xi}, u_{\rm yi}, u_{\rm zi}, u_{\rm xj}, u_{\rm yj}, u_{\rm zj} \right\}^T$ and $\mathbf{u}_{\rm r} = \left\{ \phi_{\rm xi}, \phi_{\rm yi}, \phi_{\rm zi}, \phi_{\rm xj}, \phi_{\rm yj}, \phi_{\rm zj} \right\}^T$. 
These~degrees of freedom $\mathbf{u}_{\rm t}$ and $\mathbf{u}_{\rm r}$ are used to determine displacement discontinuities \mbox{$\mathbf{u}_{\rm C} = \left\{u_{\rm n}, u_{\rm p}, u_{\rm q}\right \}^T$} at point $C$ by rigid body kinematics \citep{Kaw78} as 
\begin{equation}\label{eq:jump}
\mathbf{u}_{\rm C} = \mathbf{B}_{\rm 1} \mathbf{u}_{\rm t} + \mathbf{B}_{\rm 2} \mathbf{u}_{\rm r},
\end{equation}
where $\mathbf{B}_{1}$ and $\mathbf{B}_{2}$ are two matrices containing the rigid body information for the nodal translations and rotations, respectively, which are
\begin{equation}\label{eq:BTrans}  
\mathbf{B}_{\rm 1} =
\begin{pmatrix}
  -\mathbf{I} & \mathbf{I},
\end{pmatrix}
\end{equation}
and
\begin{equation}\label{eq:BRot}  
\mathbf{B}_{\rm 2} =
\begin{pmatrix}
  0 & -e_{\rm q} & e_{\rm p} & 0 & e_{\rm q} & -e_{\rm p}\\
  e_{\rm q} & 0 & -h/2 & -e_{\rm q} & 0 & -h/2\\
  -e_{\rm p} & h/2 & 0 & e_{\rm p} & h/2 & 0
\end{pmatrix},
\end{equation}
where $\mathbf{I}$ is a $3\times3$ unity matrix.
In matrix (\ref{eq:BRot}), $e_{\rm p}$ and $e_{\rm q}$ are the eccentricities between the midpoint of the network element and the centroid $C$ in the directions $p$ and $q$ of the local coordinate system, respectively (Figure~\ref{fig:3DTess}b).
The local coordinate system is defined by the direction $n$, which is parallel to the axis of the element, and $p$ and $q$, which are chosen as the two principal axes of the mid-cross-section. 

The displacement jump $\mathbf{u}_{\rm C}$ in Equation (\ref{eq:jump}) is transformed into strains $\boldsymbol{\varepsilon} = \left\{\varepsilon_{\rm n}, \varepsilon_{\rm p}, \varepsilon_{\rm q} \right\}^T = \mathbf{u}_{\rm C}/h$, where $h$ is the length of the structural element.
The strains are related to stresses $\boldsymbol{\sigma} = \left\{\sigma_{\rm n}, \sigma_{\rm p}, \sigma_{\rm q} \right\}^T$ by means of a material stiffness $\mathbf{D} = \left(1-\omega\right) \mathbf{D}_{\rm e}$, where $\mathbf{D}_{\rm e} = \mathrm{diag} \left\{E,\gamma E, \gamma E \right \}$. Here, $E$ is the Young's modulus and $\omega$ is the damage variable, which is further discussed in Section~\ref{sec:structuralMaterial}.
Furthermore, $\gamma$ is an input parameter, which controls Poisson's ratio of the structural network.
For $\gamma= 1$, Poisson's ratio equal to zero is obtained, which is used in this study. For this case, the structural network is elastically homogeneous under uniform modes of straining.

For the case that the global coordinate system coincides with the local one, the element stiffness matrix is
\begin{equation}
  \mathbf{K} = \dfrac{A}{h}
  \begin{pmatrix}
    \mathbf{B}_{\rm 1}^{\rm T} \mathbf{D} \mathbf{B}_{\rm 1} &  \mathbf{B}_{\rm 1}^{\rm T} \mathbf{D} \mathbf{B}_{\rm 2}\\
    \mathbf{B}_{\rm 2}^{\rm T} \mathbf{D} \mathbf{B}_{\rm 1} &  \mathbf{B}_{\rm 2}^{\rm T} \mathbf{D} \mathbf{B}_{\rm 2}
  \end{pmatrix}
  +
  \begin{pmatrix}
    \mathbf{0} & \mathbf{0}\\
    \mathbf{0} & \mathbf{B}_{\rm 1}^{\rm T} \mathbf{K_{\rm r}} \mathbf{B}_{\rm 1}         
  \end{pmatrix}
 \end{equation}

Here, $\mathbf{K}_{\rm r}$ is a matrix containing the rotational stiffness at point $C$ defined as
\begin{equation}\label{eq:Kr}
  \mathbf{K}_{\rm r} = \dfrac{(1-\omega)E}{h}
  \begin{pmatrix}
    I_{\rm p} & 0 & 0\\
    0 & I_{1} & 0\\
    0 & 0 & I_{2}
  \end{pmatrix},  
\end{equation}
where $I_{\rm p}$ is the polar moment of area, and $I_{1}$ and $I_{2}$ are the two principal second moments of area of the cross-section.
The factor $1-\omega$ in matrix (\ref{eq:Kr}) ensures that the rotational stiffness reduces to zero for a fully damaged cross-section ($\omega=1$).
For an elastic constitutive model, the present structural element is identical to the one described in \citet{BerBol06}.
All geometrical information of the network element is contained in the element formulation. In this way, the constitutive model relating stresses to strains depends only on properties of the material.
This structure is preferred over one that incorporates geometrical information in the constitutive model, since it facilitates the adoption of constitutive modelling frameworks that are commonly used for continuum approaches.

\subsubsection{Structural Material} \label{sec:structuralMaterial}
The inelastic structural response of the material during fracture is described by a scalar damage model \citep{MazPij89} of the form
\begin{equation}\label{eq:stressStrainLaw}
 \boldsymbol{\sigma} = \left(1-\omega\right) \mathbf{D}_{\rm e} \boldsymbol{\varepsilon}.
\end{equation} 

The damage variable $\omega$ is a function of the history variable $\kappa_{\rm d}$ \citep{Lemaitre90}, which is, in turn, determined by the loading function
\begin{equation}
f_{\rm d}(\boldsymbol{\varepsilon},\kappa_{\rm d}) = \varepsilon_{\rm eq} \left( \boldsymbol{\varepsilon} \right) - \kappa_{\rm d}, 
\end{equation}
and the loading--unloading conditions 
\begin{equation}\label{loadunload}
f_{\rm d} \leq 0 \mbox{,} \hspace{0.5cm} \dot{\kappa}_{\rm d} \geq 0 \mbox{,} \hspace{0.5cm} \dot{\kappa}_{\rm d} f_{\rm d} = 0.
\end{equation}

The equivalent strain
\begin{equation} \label{eq:equiv}
\varepsilon_{\rm eq}(\varepsilon_{\rm n},\varepsilon_{\rm p}, \varepsilon_{\rm q}) = \dfrac{1}{2} \varepsilon_0 \left( 1-c_{\rm s} \right) + \sqrt{\left( \dfrac{1}{2} \varepsilon_0 (c_{\rm s}-1) + \varepsilon_{\rm n}\right)^2 + \dfrac{ c_{\rm s} \gamma^2 \left (\varepsilon_{\rm p}+ \varepsilon_{\rm q}\right)^2}{q_{\rm s}^2}}
\end{equation} 
corresponds to an ellipsoidal envelope in the stress space.
For pure tensile loading, the stress is limited by the tensile strength $f_{\rm t} = E \varepsilon_{0}$.
For pure shear and pure compression, the stress is limited by the shear strength $f_{\rm q} = q_{\rm s} f_{\rm t}$ and the compressive strength $f_{\rm c} = c_{\rm s} f_{\rm t}$, respectively.

The damage function is determined by using an exponential stress-crack law in pure tension of the form 
\begin{equation} \label{eq:exp}
\sigma_{\rm n} = f_{\rm t} \exp \left(-\dfrac{w_{\rm n}}{w_{\rm f}}\right),
\end{equation}
where $w_{\rm n} = \omega h \varepsilon_{\rm n}$ is the crack opening under monotonic tension and $\varepsilon_{\rm n}$ is the tensile strain.
This crack opening is the first component of the crack opening vector $\mathbf{w} = \omega h \boldsymbol{\varepsilon}$, which is used for the coupling of the structural and mass transport model.
The normal stress in Equation (\ref{eq:exp}) is also expressed in terms of the stress-strain law in Equation (\ref{eq:stressStrainLaw}) as
\begin{equation}\label{eq:uni}
\sigma_{\rm n} =  \left(1-\omega \right) E \varepsilon_{\rm n}.
\end{equation}
Comparing the right-hand sides of Equations (\ref{eq:exp})~and~(\ref{eq:uni}), and replacing $\varepsilon_{\rm n}$ by $\kappa_{\rm d}$, since a monotonically increasing tensile strain is assumed, the nonlinear equation
\begin{equation}
\left(1-\omega \right) E \kappa_{\rm d} = f_{\rm t} \exp \left(-\dfrac{\omega h \kappa_{\rm d}}{w_{\rm f}}\right)
\end{equation}
is obtained from which the damage parameter $\omega$ is determined iteratively using a Newton method.
In Equation (\ref{eq:exp}), parameter $w_{\rm f}$ determines the initial slope of the softening curve and is related to the fracture energy as $G_{\rm F} = f_{\rm t} w_{\rm f}$.
The input parameters for the structural part of the model are the Young's modulus $E$, the tensile strength $f_{\rm t}$, fracture energy $G_{\rm F}$, shear strength $f_{\rm q}$ and compressive strength $f_{\rm c}$.
These input parameters can be determined from inverse analysis of elementary structural tests of the specific geomeaterial of interest.

\subsection{Transport Model} 
\label{sec:transElem}
For the transport part of the model, a 3D network of 1D transport elements is used to discretise the nonstationary transport equation \citep{MaeIshKis08}
\begin{equation} \label{eq:flow}
\dfrac{\partial P_{\rm c}}{\partial t} - \mathrm{div} \left(\alpha \mathrm{grad} P_{\rm c} \right) = 0,
\end{equation}
subject to
\begin{equation}
P_{\rm c} = g\left( \mathbf{x} \right) \hskip 1cm \mbox{on $\Gamma_1$}, 
\end{equation}
and
\begin{equation}
f = - \dfrac{\partial P_{\rm c}}{\partial \mathbf{n}} \hskip 1cm \mbox{on $\Gamma_2$}, 
\end{equation}
where $P_{\rm c}$ is the capillary suction, $t$ is the time, $\alpha$ is the conductivity, $f$ is the outward flux normal to the boundary ($\mathbf{n}$-direction) and $\mathbf{x}$ is the position in the domain $\Omega$. Furthermore, $\Gamma_1$ and $\Gamma_2$ are the boundary segments with prescribed suction and flux, respectively. The capillary suction $P_{\rm c}$ in an unsaturated material is defined as $P_{\rm c} = P_{\rm d}-P_{\rm w}$, where $P_{\rm d}$ is the pressure in the drying fluid and $P_{\rm w}$ is the pressure in the wetting fluid. Here, $P_{\rm d}$ is assumed to be zero, which is a common assumption for modelling the water retention in unsaturated materials subjected to ambient temperatures \citep{MaeIshKis08}.

\subsubsection{Transport Element}
\begin{figure}[ht]
\centering
    \includegraphics[width=4cm]{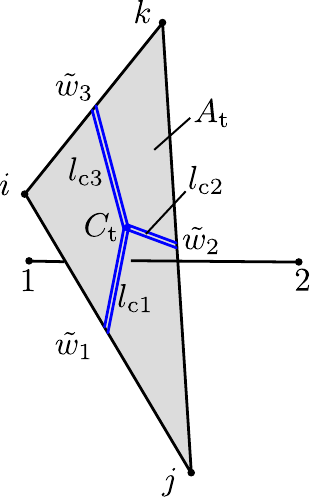}
  \caption{Influence of cracking on transport.}
\label{fig:transport3dCrack}
\end{figure}

The discrete form of Equation (\ref{eq:flow}) for a 1D transport element shown in Figure~\ref{fig:3DTess}c is
\begin{equation}\label{eq:flowDiscrete}
  \mathbf{C}_{\rm e} \dfrac{\partial \boldsymbol{P}_{\rm c}}{\partial t} - \boldsymbol{\alpha}_{\rm e} \boldsymbol{P}_{\rm{c}} = \mathbf{f},
\end{equation}
where $\boldsymbol{\alpha}_{\rm e}$ and $\mathbf{C}_{\rm e}$ are the 1D element conductivity and capacity matrices, respectively, and $\mathbf{f}$ are the external fluxes \citep{LewMorTho96,BolBer04}. The degrees of freedom of the transport elements are the capillary suction $\boldsymbol{P}_{\rm c} = \left(P_{\rm c1}, P_{\rm c2}\right)^T$. Within the context of a one-dimensional finite element formulation \citep{LewMorTho96}, Galerkin's method is used to construct the elemental capacity matrix $\mathbf{C}_{\rm e}$ as
\begin{equation}\label{eq:capacity}
\mathbf{C}_{\rm e} = c \dfrac{A_{\rm t} h_{\rm t}}{12}
\begin{pmatrix}
2 & 1\\
1 & 2
\end{pmatrix},
\end{equation}
where $c$ is the capacity of the material, $A_{\rm t}$ is the cross-sectional area of the tetrahedron face associated with the transport element (Figure~\ref{fig:transport3dCrack}), and $h_{\rm t}$ is the length of the transport element.

Likewise, based on Galerkin's method \citep{LewMorTho96}, the elemental conductivity matrix is defined as
\begin{equation}\label{eq:conMat}
\boldsymbol{\alpha}_{\rm e} = \dfrac{A_{\rm t}}{h_{\rm t}} \alpha
\begin{pmatrix} 
1 & -1\\
-1 & 1
\end{pmatrix},
\end{equation}
where $\alpha$ is the conductivity of the material, which is the sum of two components
\begin{equation} \label{eq:conSplit}
\alpha = \alpha_0 + \alpha_{\rm c},
\end{equation}
where $\alpha_0$ is the initial conductivity of the undamaged material and $\alpha_{\rm c}$ is the change of the conductivity due to fracture.

\subsubsection{Transport Material}\label{sec:transMaterial}
In the present example, the network approach is applied to mass transport in a general unsaturated geomaterial using techniques introduced originally by van Genuchten for soils \citep{Gen80}, but also applied to other geomaterials, such as cementitious materials \cite{BarMaiLas99}.
According to van Genuchten in \citep{Gen80}, the conductivity of the undamaged material $\alpha_0$ is defined as
\begin{equation}
\alpha_0 = \dfrac{\rho \kappa}{\mu} \kappa_{\rm r} \left(S\right),
\end{equation}
where $\rho$ is the density of the fluid, $\mu$ is the dynamic (absolute) viscosity, $\kappa$ is the intrinsic permeability and $\kappa_{\rm r}$ is the relative permeability as a function of the degree of saturation.
This degree of saturation is defined as
\begin{equation}\label{eq:saturation}
S = \dfrac{\theta -\theta_{\rm r}}{\theta_{\rm s} - \theta_{\rm r}},
\end{equation}
with the moisture content $\theta$, the residual moisture content $\theta_{\rm r}$ and the saturated moisture content $\theta_{\rm s}$ of the specific geomaterial \citep{Gen80}.
Furthermore, the relative permeability $\kappa_{\rm r}$ is
\begin{equation}\label{eq:relCon}
\kappa_{\rm r} \left(S\right) = \sqrt{S}\left(1-\left(1-S^{1/m}\right)^{m}\right)^2,
\end{equation}
where $m$ is a model parameter \citep{Gen80}.
The saturation is related to the capillary suction as
\begin{equation}\label{eq:sat}
S\left(P_{\rm c}\right) = \left(1+\left(\dfrac{P_{\rm c}}{a}\right)^{\dfrac{1}{1-m}}\right)^{-m}
\end{equation}
where $a$ is another model parameter. Physical justification of parameters $m$ and $a$ in Equations (\ref{eq:relCon}) and (\ref{eq:sat}) are given by van Genuchten \citep{Gen80}.
The second term in Equation (\ref{eq:conSplit}) describes the influence of fracture on conductivity.
It is defined as
\begin{equation}\label{eq:crackDiff}
\alpha_{\rm c} = \xi \dfrac{\rho}{12 \mu A_{\rm t}} \sum_{i=1}^{3} \tilde{w}_{i}^3 l_{\rm{c}i},
\end{equation}
where $\tilde{w}_{i}$ and $l_{\rm ci}$ are the equivalent crack openings and crack lengths (see Figure~\ref{fig:transport3dCrack}) of neighbouring structural elements, which are located on the edges of the cross-section, and $\xi$ is a tortuosity factor. For mortars, crack tortuosity considered by $\xi$ may reduce flow by a factor of 4 to 6, relative to that between smooth parallel plates \citep{AkhShaRaj12}.
Here, $\tilde{w} = \left|\mathbf{w}\right|$ is the magnitude of the crack opening $w$ defined in Section~\ref{sec:structuralMaterial}. 
The relation in Equation (\ref{eq:crackDiff}) expresses the well known cubic law, which has shown to produce good results for transport in fractured geomaterials \cite{WitWanIwaGal80}. In Equation (\ref{eq:crackDiff}), $w_{\rm i}$ is assumed to act over $l_{\rm ci}$ (i.e.,~the equivalent crack opening is uniform over the element crack length). The approach adequately represents variations in opening along the crack trajectory, provided the mesh is sufficiently fine.

The way that the crack openings in the structural elements influence the conductivity of a transport element is schematically shown in Figure~\ref{fig:transport3dCrack}.
For instance, for the transport element $o$--$p$, three structural elements ($i$--$k$,
 $k$--$j$ and $i$--$j$) bound the cross-section of the transport element.
Thus, the conductivity will be influenced by these three elements according to Equation (\ref{eq:crackDiff}) in proportion to their equivalent crack widths and the crack lengths. This crack length (shown by blue double lines in Figure~\ref{fig:transport3dCrack}) is defined as the length from the midpoint of the structural element to the centroid $C_t$ of the transport element cross-section.

The capacity $c$ in Equation (\ref{eq:capacity}) is defined as $c = - \rho \partial \theta/\partial P_{\rm c}$. Using Equation (\ref{eq:saturation}), this expression can be written as
\begin{equation}
c = -\rho \dfrac{\partial S}{\partial P_{\rm c}} \left(\theta_{\rm s} - \theta_{\rm r}\right).
\end{equation}

It is assumed that $c$ is independent of the cracking described by the structural part.
The input parameters of the transport part are the density $\rho$ and dynamic viscosity $\mu$ of the wetting fluid, the permeability of the saturated uncracked material $\kappa$, the saturated and residual wetting fluid content, $\theta_{\rm s},$ and $\theta_{\rm r}$, respectively.
Furthermore, parameters $m$ and $a$ of the van Genuchten constitutive model, and the tortuosity parameter $\xi$ are needed.

The structural network is adept at simulating fracture in multi-phase representations of concrete, in which the matrix, aggregates, and matrix-aggregate interfaces are explicitly represented \citep{GraJir10,AsaLanBol11}. Study of the influence of interface fracture on effective permeability is one potential application of the proposed dual-network approach.

\section{Analyses}
In the proposed coupled network approach, the transport elements, which describe both the transport through continuum and fractures, are placed on the edges of the Voronoi polyhedra.
This differs from the commonly used approach in which the elements are located at the edges of the Delaunay tetrahedra \citep{BolBer04}.
The performance of this new approach is investigated by three benchmark~tests. The numerical analyses are performed with OOFEM, an open-source object-oriented finite element program \citep{Pat12} extended by the present authors.

\subsection{Steady-State Potential Flow}\label{sec:steadyFlow}
\begin{figure}[ht]
\centering
\includegraphics[width=0.9\textwidth]{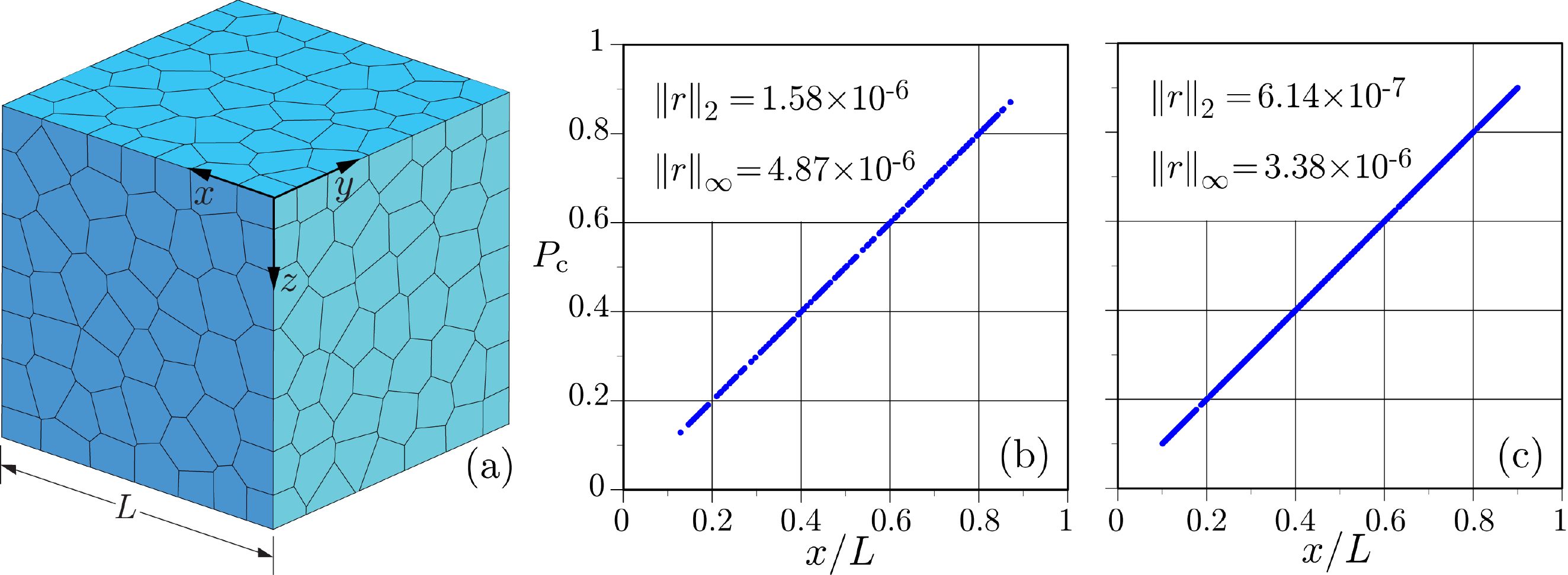}
\caption{Steady-state simulation of potential flow: (\textbf{a}) Voronoi tessellation of domain; (\textbf{b}) conventional network solution; and (\textbf{c}) proposed network solution.}\label{fig:ssSolution}
\end{figure}

For the first benchmark, a homogeneous material is discretised as shown in Figure~\ref{fig:ssSolution}a.
  The Delaunay/Voronoi discretisation of the domain is based on a set of randomly inserted nodes.
  Table \ref{tab:featureCount} compares the numbers of nodes/elements forming both network types depicted in Figure~\ref{fig:2D}:
the conventional approach (in which transport elements are on the Delaunay edges) and the proposed approach (in which transport elements are on the Voronoi edges).  It is clear that the proposed approach is computationally more expensive. 
\begin{table}[ht]
\centering
\caption{Network feature counts.\vspace{-6pt}}
\label{tab:featureCount}
\begin{tabular}{ccccc}
\hline
\textbf{Network Type}& \textbf{Node Definition} & \textbf{Element Definition} & \textbf{Nodal Count} $^*$ & \textbf{Element Count} $^*$ \\
\hline
Conventional & Delaunay vertex & Delaunay edge & 330 & 1800 \\
Proposed      & Voronoi vertex     &  Voronoi edge & 2880 & 5440 \\
\hline
\end{tabular}\\
\begin{tabular}{cccc}
\multicolumn{1}{c}{\footnotesize $^*$ rounded to nearest ten.}
\end{tabular}
\end{table}
The material is subjected to a pressure difference between the $x$-faces of the domain: $P_{\rm c} (x \! = \! 0) \! = \! 0$ and $P_{\rm c} (x \! = \! L) \! = \! 1$.
For this test, a special case of the constitutive model presented in Section~\ref{sec:transMaterial} has been used by assuming the conductivity and capacity to be constant with values of $\alpha = c = 1$.
Both~networks accurately represent the steady-state solution, as shown by the nodal potentials plotted in Figure~\ref{fig:ssSolution}. Pressure values are not plotted for the nodes associated with prescribed boundary conditions. The discrete error norms presented in the figures are:
\begin{equation}
||r||_\infty =  \max_{m=1,...,M} |r_m|,
\end{equation}
\begin{equation}
 ||r||_2 = \left( \frac{1}{M} \sum_{m=1}^M |r_m|^2 \right)^{1/2},
\end{equation}
where $r_m = P_{\rm c}(\mathbf{x}_m) - P_{\rm ch}({\mathbf x}_m)$ is the difference between the theoretical and numerical solutions, respectively, at the position of node $m$; and $M$ is the number of unconstrained nodal points.

\subsection{Nonstationary Transport Analysis}\label{sec:transBench}
\begin{figure}[ht]
\centering
 \includegraphics[width=10cm]{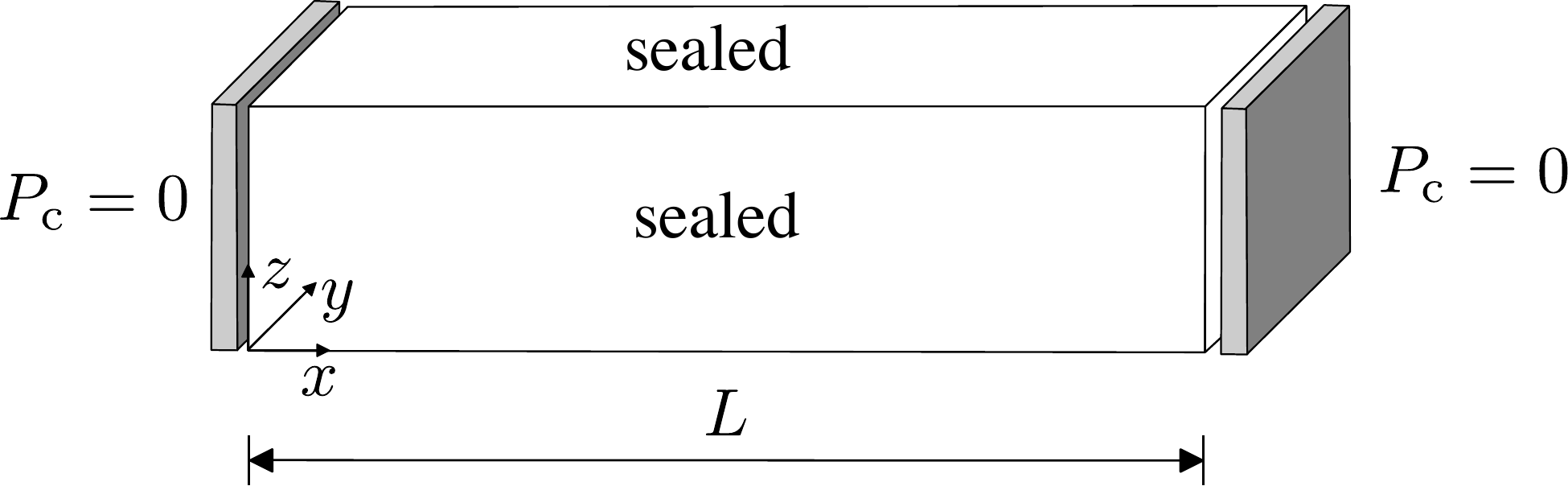}
\caption{Geometry and boundary conditions for the nonstationary transport benchmark.}
\label{fig:transportGeometry}
\end{figure}

\begin{figure}[ht]
\centering
  \includegraphics[width=9cm]{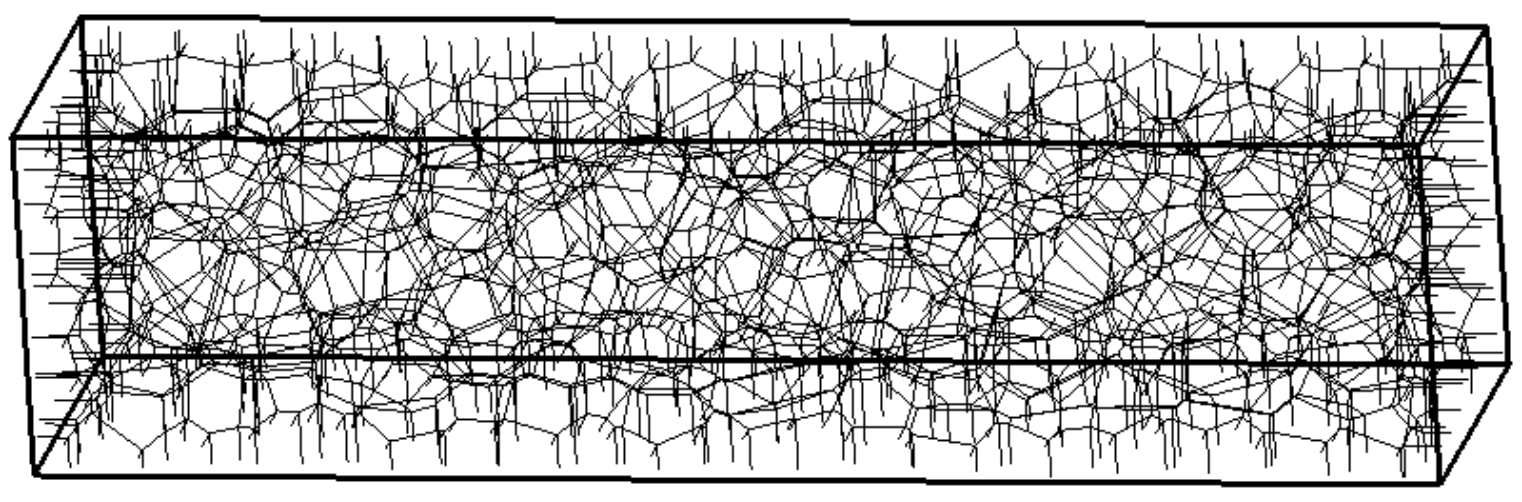}
\caption{Coarse network for the transport analysis.}
\label{fig:transportMesh}
\end{figure}

For the second benchmark, nonstationary mass transport through undamaged material was studied.
The geometry and boundary conditions are shown in Figure~\ref{fig:transportGeometry}.
The two ends of the specimen are subjected to zero pressure whereas all other boundaries are considered to be sealed. 
For this test, again the special case of $\alpha = c = 1$ for the constitutive model presented in Section~\ref{sec:transMaterial} has been used.
The initial condition at all nodes is $P_{\rm c}\left(x,t\right) = P_0 \sin \left(\dfrac{\pi x}{L}\right)$.
This assumption allows for a comparison with the analytical solution $P_{\rm c} = P_0 \sin\left(\dfrac{\pi x}{L}\right) \exp \left(-\dfrac{\pi^2}{L^2} t\right)$ reported in \cite{BolBer04}.
Three transport networks with minimum distances between Delaunay vertices of \mbox{$d_{\rm min}/L = 0.06$},~$0.045$~and~$0.03$ are used.
The coarse network with $d_{\rm min}/L = 0.06$ is shown in Figure~\ref{fig:transportMesh}.

\begin{figure}[ht]
\centering
    \includegraphics[width=9cm]{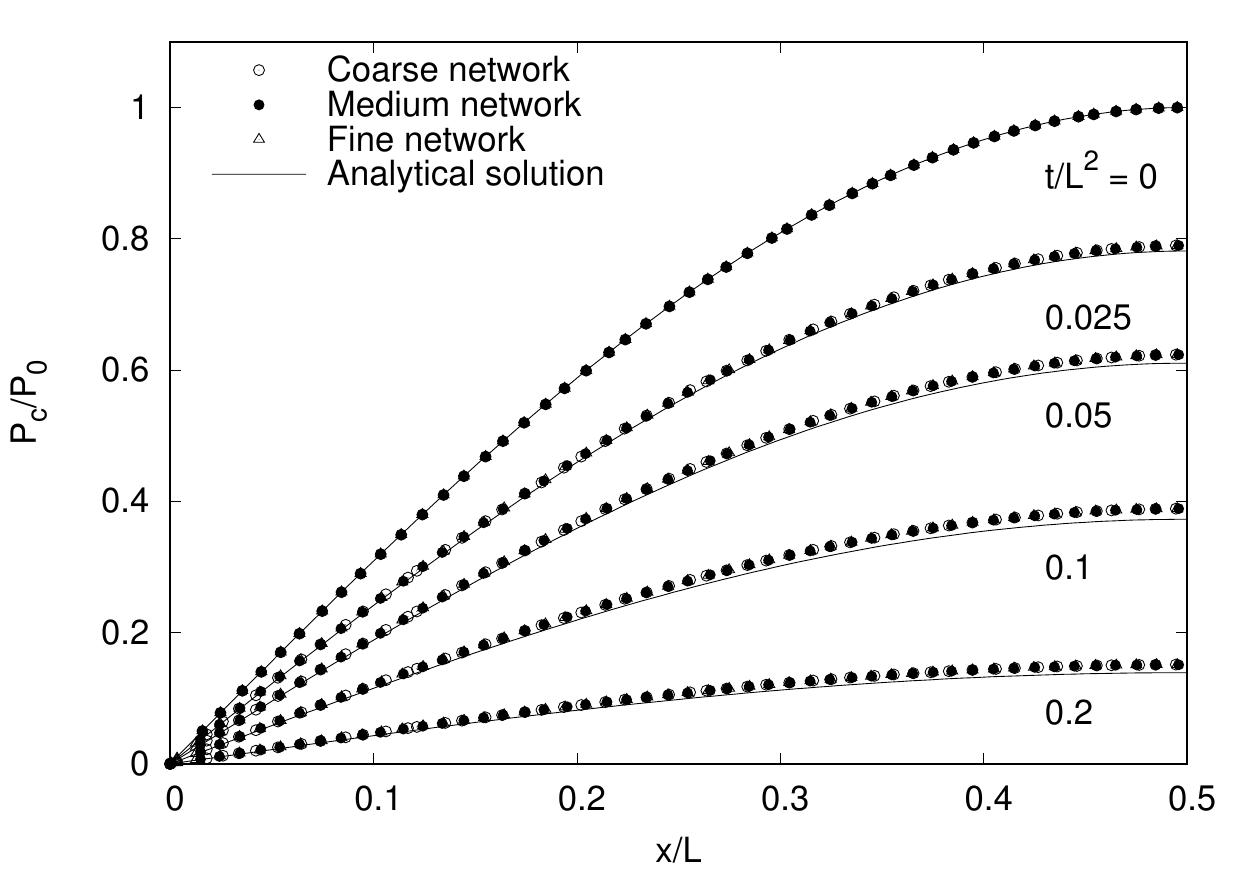}
\caption{Capillary suction distributions due to non-stationary transport.}  
\label{fig:compTransport}
\end{figure}

The capillary suction distributions for four time steps and the three different network sizes are shown in Figure~\ref{fig:compTransport} for a symmetric portion of the model.
For comparing the network results with the analytical results, the vertices were divided into groups with respect to their $x$-coordinate. For each group of vertices the mean of the $x$-coordinate and capillary suction are presented.
The Voronoi-edge based network agrees well with the analytical solution of the capillary suction distribution without exhibiting any dependence on the element size.
Any differences between the numerical and analytical solution originate from the time discretisation, rather than the new spatial discretisation.

\subsection{Coupled Structural-Transport Benchmark}\label{sec:strucBench}
\begin{figure}[ht]
\centering
\begin{tabular}{c}
  \includegraphics[width=9cm]{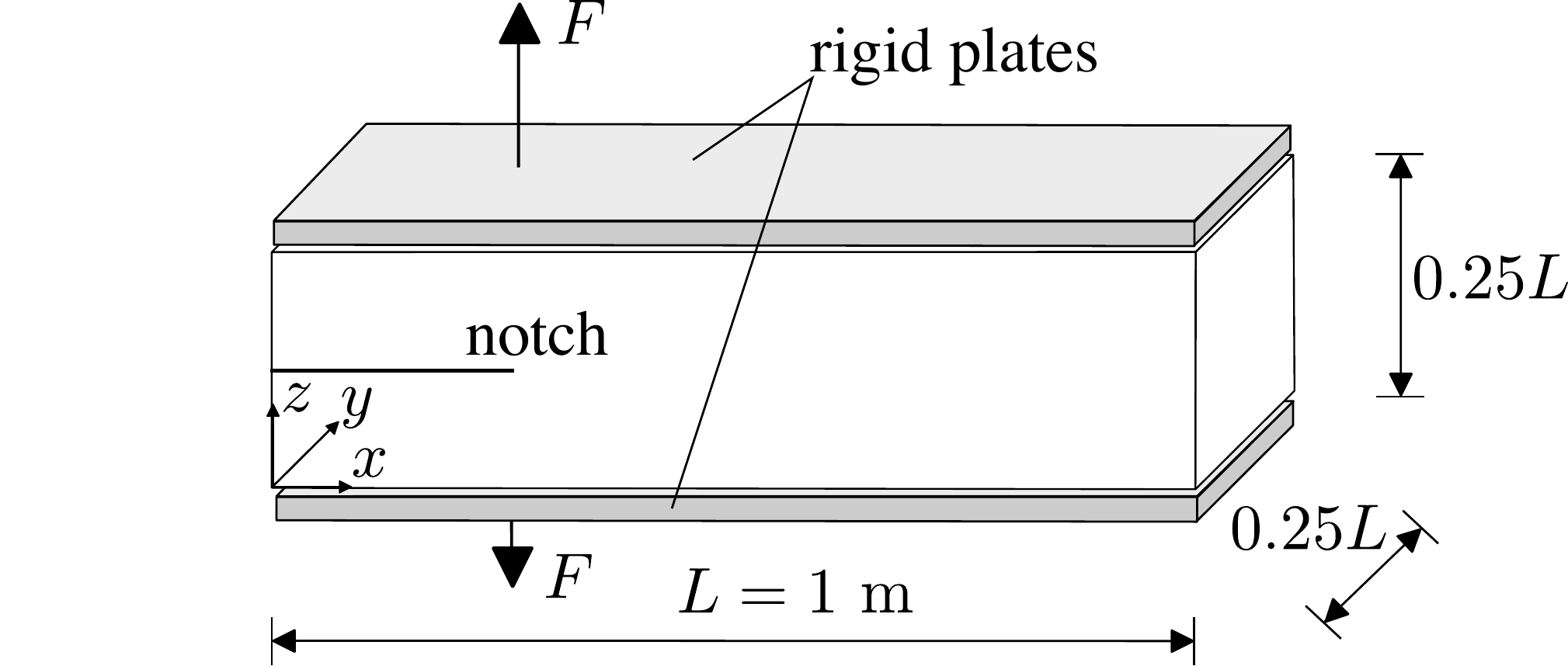} \\
  (\textbf{a}) \vspace{0.5cm}\\
  \includegraphics[width=9cm]{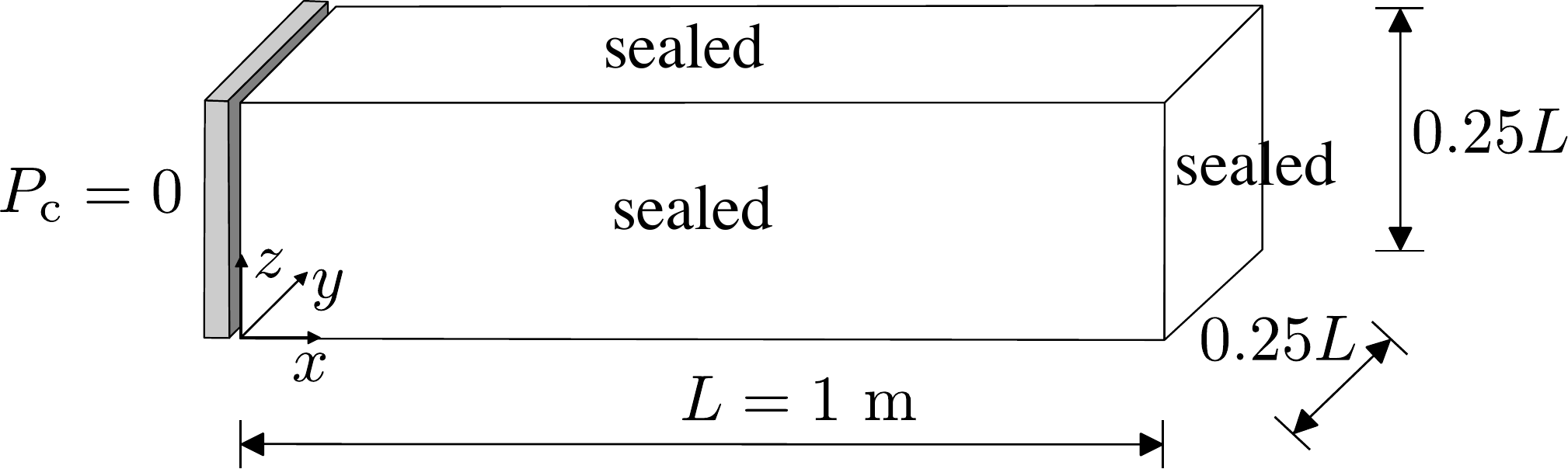} \\
  (\textbf{b})
\end{tabular}
\caption{Geometry and loading setup of the structural and transport benchmark.}
\label{fig:structuralGeometry}
\end{figure}

In the third benchmark, the structural and transport models are coupled.
Firstly, a double cantilever beam is used to assess the capability of the structural model to describe fracture without any pathological network dependence.
Then, fluid transport through the fractured specimen at an intermediate loading stage of the structural analysis is modelled for different networks with different element sizes.
The geometry and loading setup for the structural and transport tests are shown in Figure~\ref{fig:structuralGeometry}a,b, respectively. For the structural analysis, the load is applied at $x=0.25L$.

\begin{figure}[ht]
\centering
\begin{tabular}{c}
    \includegraphics[width=10cm]{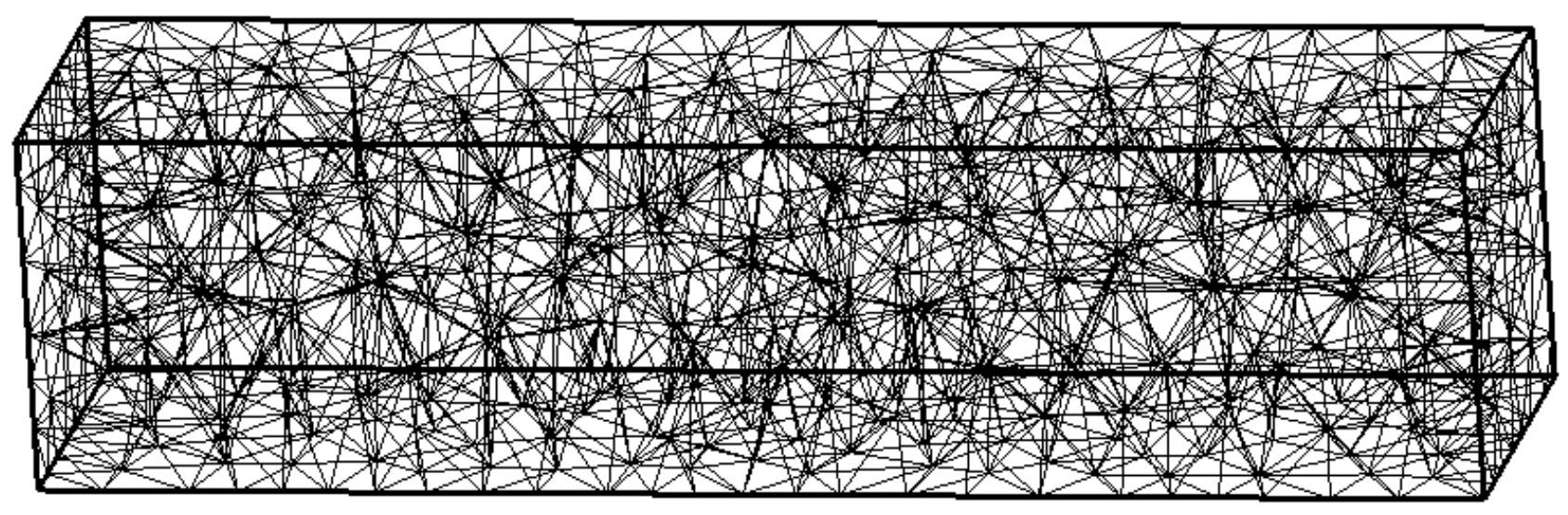}\\
    (\textbf{a})\\
    \includegraphics[width=10cm]{./figTransportMesh.pdf}\\
    (\textbf{b})
\end{tabular}
  \caption{The coarse dual networks ($d_{\rm min}/L = 0.06$) for (\textbf{a}) structural and (\textbf{b}) transport analysis.}
  \label{fig:structuralMesh}
\end{figure}

For the transport component of the analyses, three networks with minimum distances between Delaunay vertices of $d_{\rm min}/L = 0.06$,~$0.045$~and~$0.03$ were used.
The structural and transport networks with $d_{\rm min}/L = 0.06$ are shown in Figure~\ref{fig:structuralMesh}a,b, respectively. As noted earlier, transport elements local to the boundaries are perpendicular to the specimen surfaces.
The input parameters for the structural constitutive model are $E = 30$~GPa, $f_{\rm t} = 3$~MPa and $G_{\rm F} = 120$~N/m, which are representative of concrete materials.
A notch of length $0.25L$ is introduced by reducing the tensile strength $f_{\rm t}$ of elements crossing the notch to 1\% of the original value.

\begin{figure}[ht]
\centering
  \includegraphics[width=12cm]{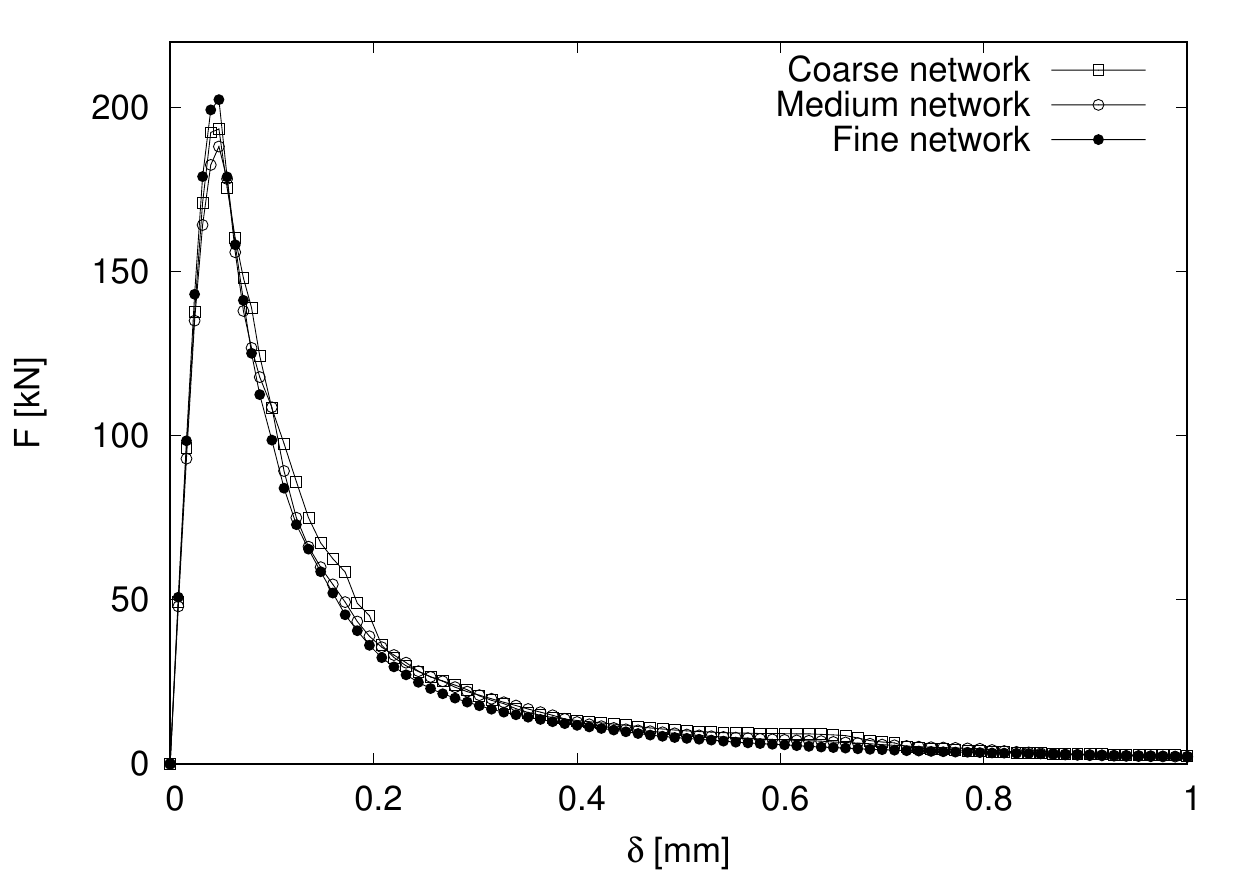}
\caption{Load versus load-point-displacement results for three networks.}
\label{fig:ldMesh}
\end{figure}

\begin{figure}[ht]
\centering
    \begin{tabular}{cc}
(\textbf{a}) & \includegraphics[width=10cm]{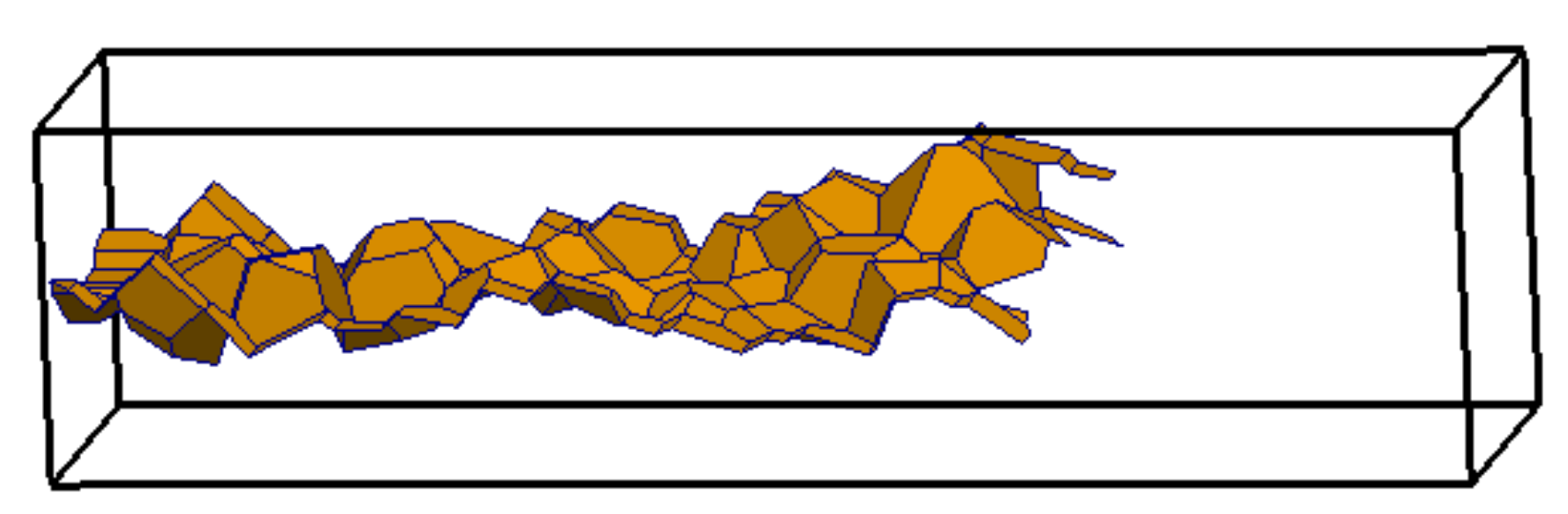}\\
(\textbf{b}) & \includegraphics[width=10cm]{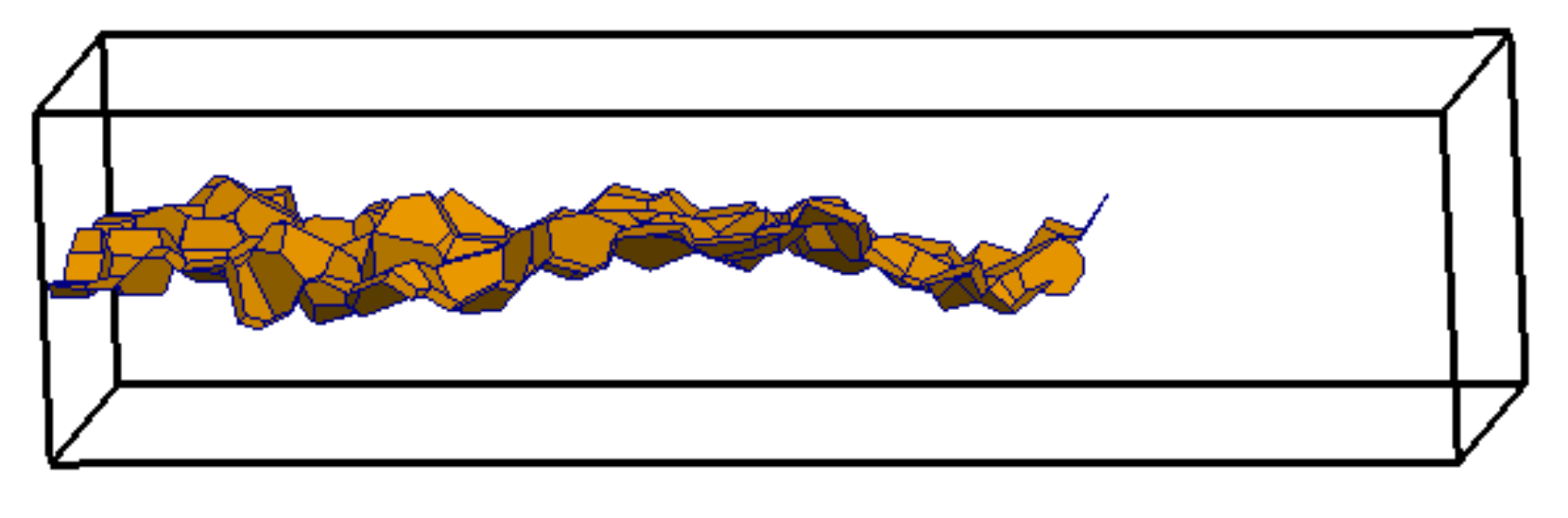}\\
(\textbf{c}) & \includegraphics[width=10cm]{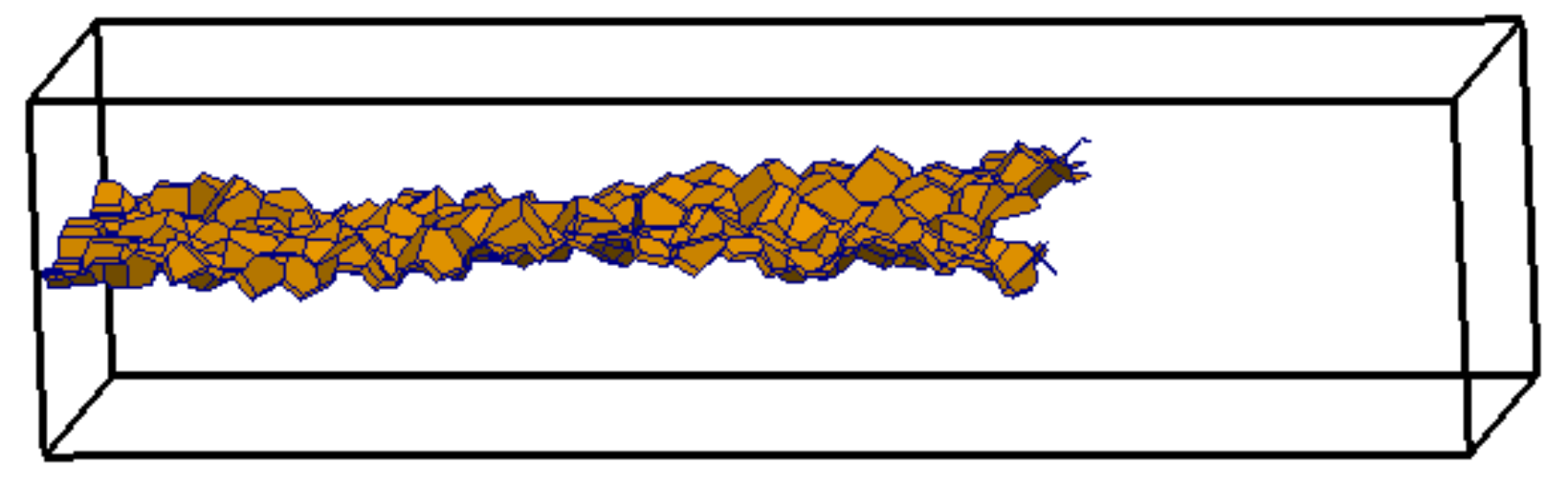}
  \end{tabular}
\caption{Crack patterns for (\textbf{a}) coarse; (\textbf{b}) medium; and (\textbf{c}) fine network for a load-point-displacement of $\delta = 0.15$~mm in Figure~\ref{fig:ldMesh}. The shaded polygons represent the mid-cross-sections of elements with $\tilde{w} > 10$~$\mu$m.}
\label{fig:crackMesh}
\end{figure}

The load-displacement curves from the structural benchmark for the three networks are shown in Figure~\ref{fig:ldMesh}.
There is little difference between the responses obtained with the three networks.
Fracture is indicated by shading the mid-cross-sections of elements in which the equivalent crack opening has reached a threshold value.
The mid-cross-sections of elements with damage corresponding to an equivalent crack opening $\tilde{w} > 10$~$\mu$m are shown in Figure~\ref{fig:crackMesh} for the three different networks at a load-point-displacement of $\delta=0.15$~mm in Figure~\ref{fig:ldMesh}.

The transport network uses the same geometry as in the nonstationary transport test in Section~\ref{sec:transBench}.
However, the boundary and initial conditions, and the material input parameters, are changed so that the influence of fracture could be studied more effectively.
On the left-hand side of the model, the boundary is subjected to $P_{\rm c} = 0$. Furthermore, the initial capillary suction of all other nodes is set to $P_{\rm c} = 1.736$~MPa, which for the chosen material parameters corresponds to an initial saturation of $S_{\rm init}= 0.5$. 
Other input parameters for the transport problem are: $\alpha_0 = 1\times 10^{-17}$~m$^2$, $\theta_{\rm s} = 0.1$, $\theta_{\rm r} = 0$, $a = 1$~MPa, $m=0.5$ and $\xi = 0.001$. The transport analysis is performed for crack patterns obtained at a displacement of $\delta = 0.15$~mm in Figure~\ref{fig:ldMesh}.  

\begin{figure}[ht]
\centering
      \includegraphics[width=10cm]{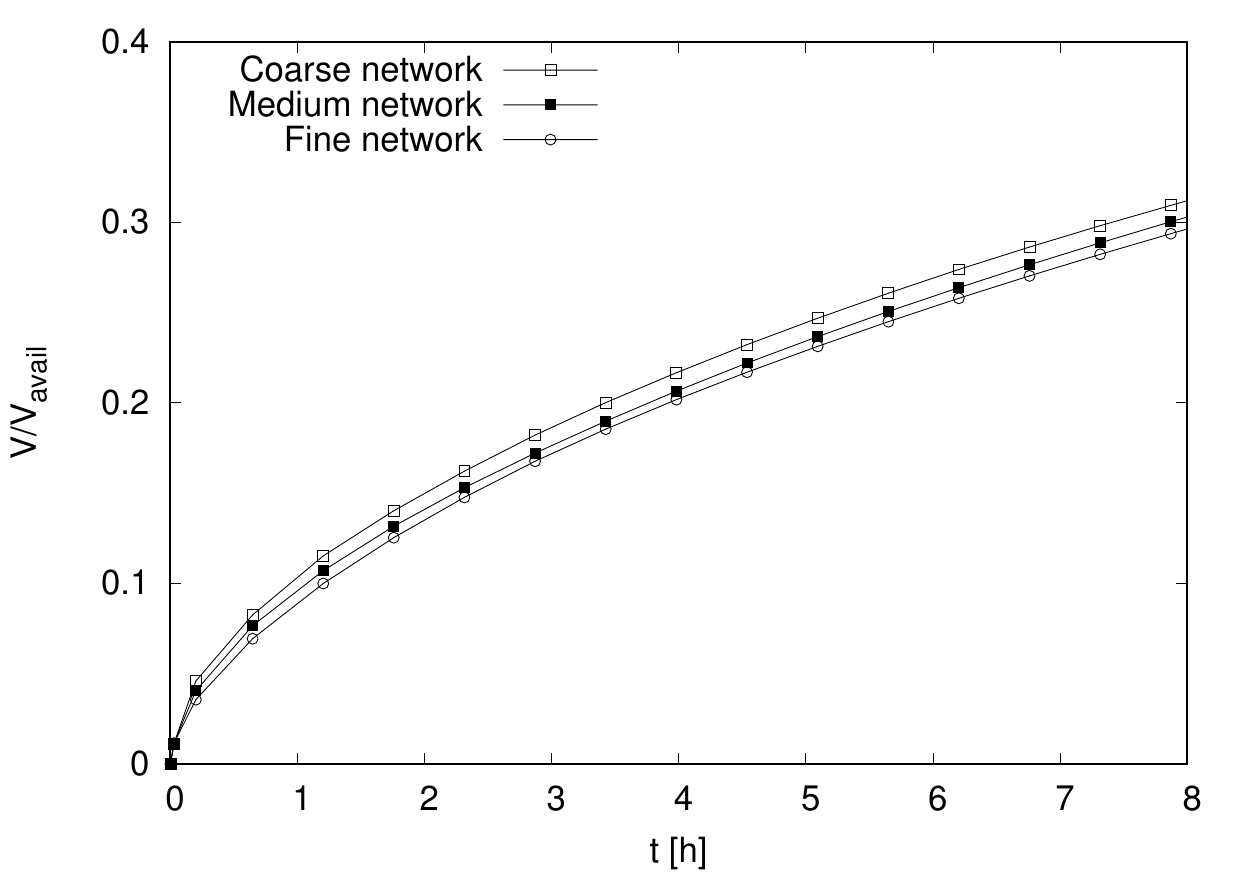}
\caption{Influence of element size on the cumulative volume of inflow normalised by the domain~volume.}
\label{fig:inflow}
\end{figure}

Results for the cumulative volume of inflow at the left side of the specimen normalised by the available volume to be filled, from the time of initial wetting, are presented in Figure~\ref{fig:inflow}.
The available volume to be filled is $V_{\rm avail} = \left(1-S_{\rm init}\right) \theta_{\rm s} V_{\rm tot}$, where $V_{\rm tot} = L \times 0.25 L \times 0.25 L$ is the total volume of the specimen.
The inflow is practically independent of the element size.

\begin{figure}[ht]
\centering
    \begin{tabular}{cc}
      \includegraphics[width=11cm]{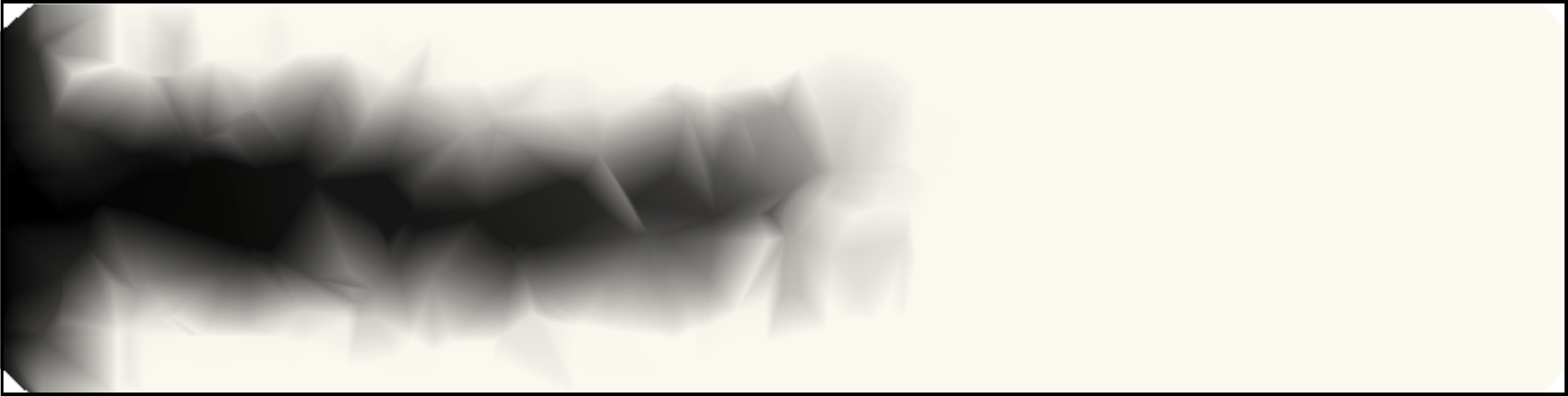} & \includegraphics[width=3cm]{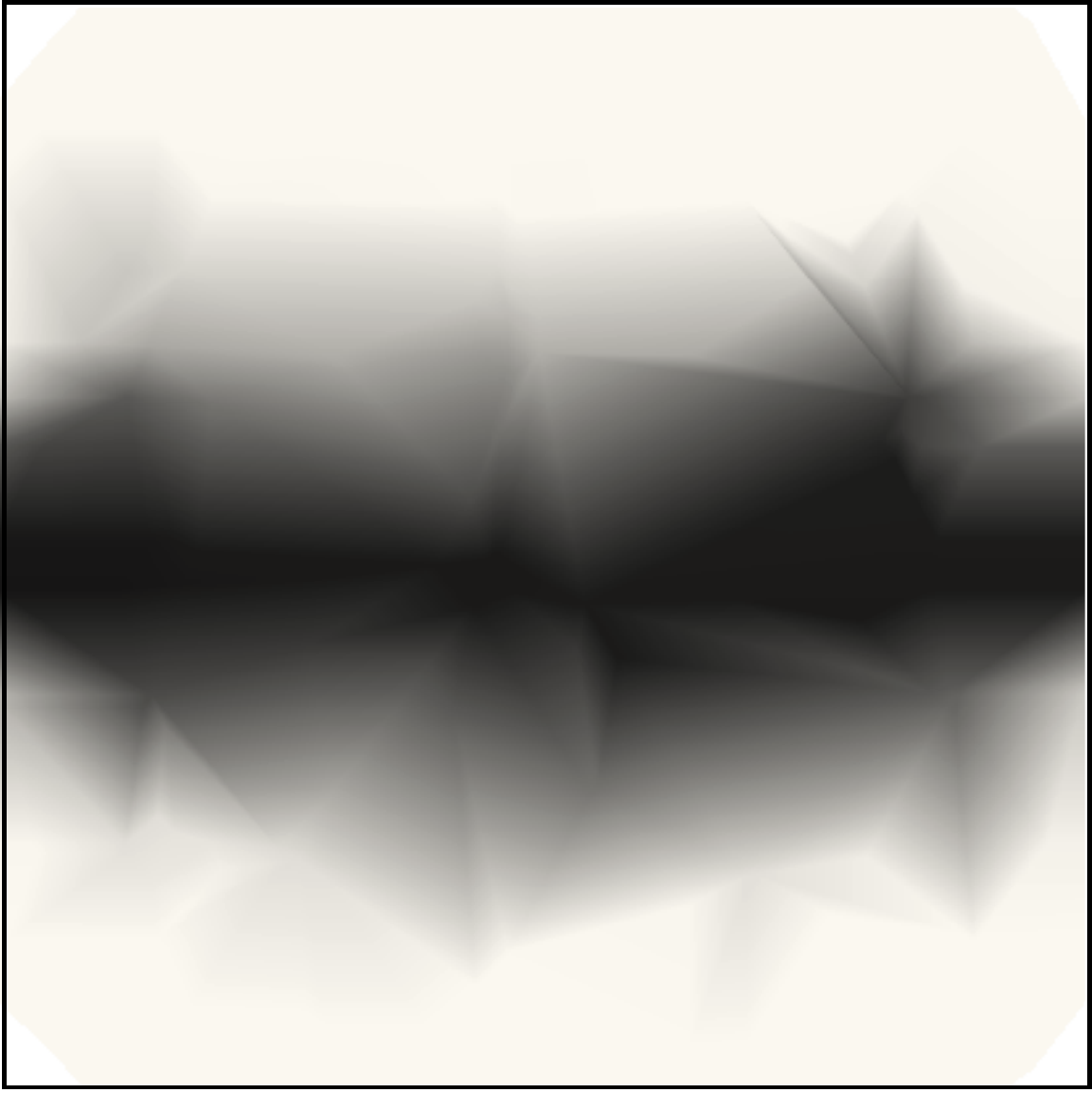}\\
     coarse\\
     \includegraphics[width=11cm]{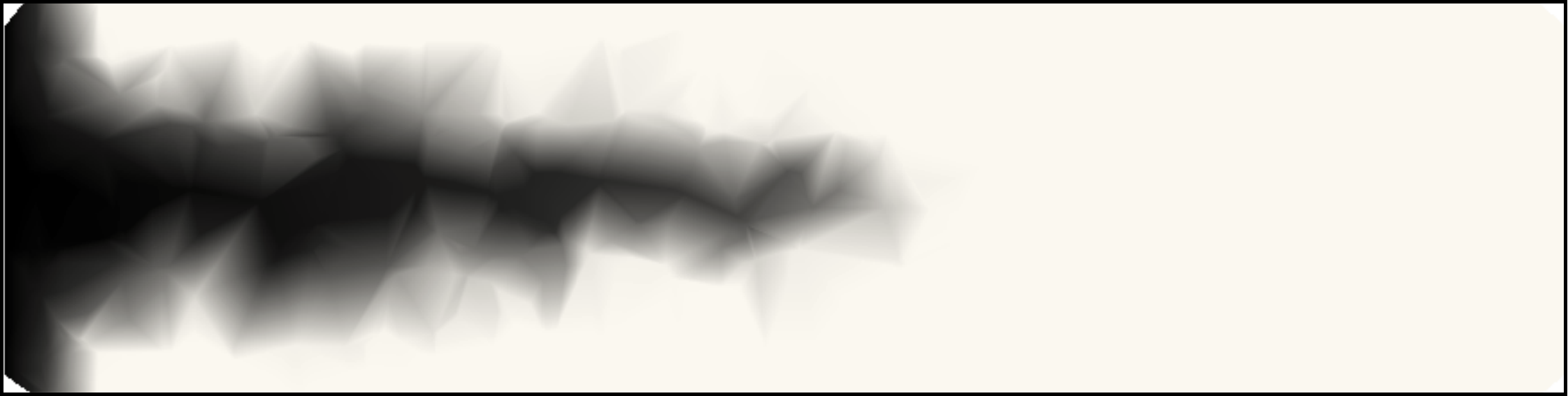} & \includegraphics[width=3cm]{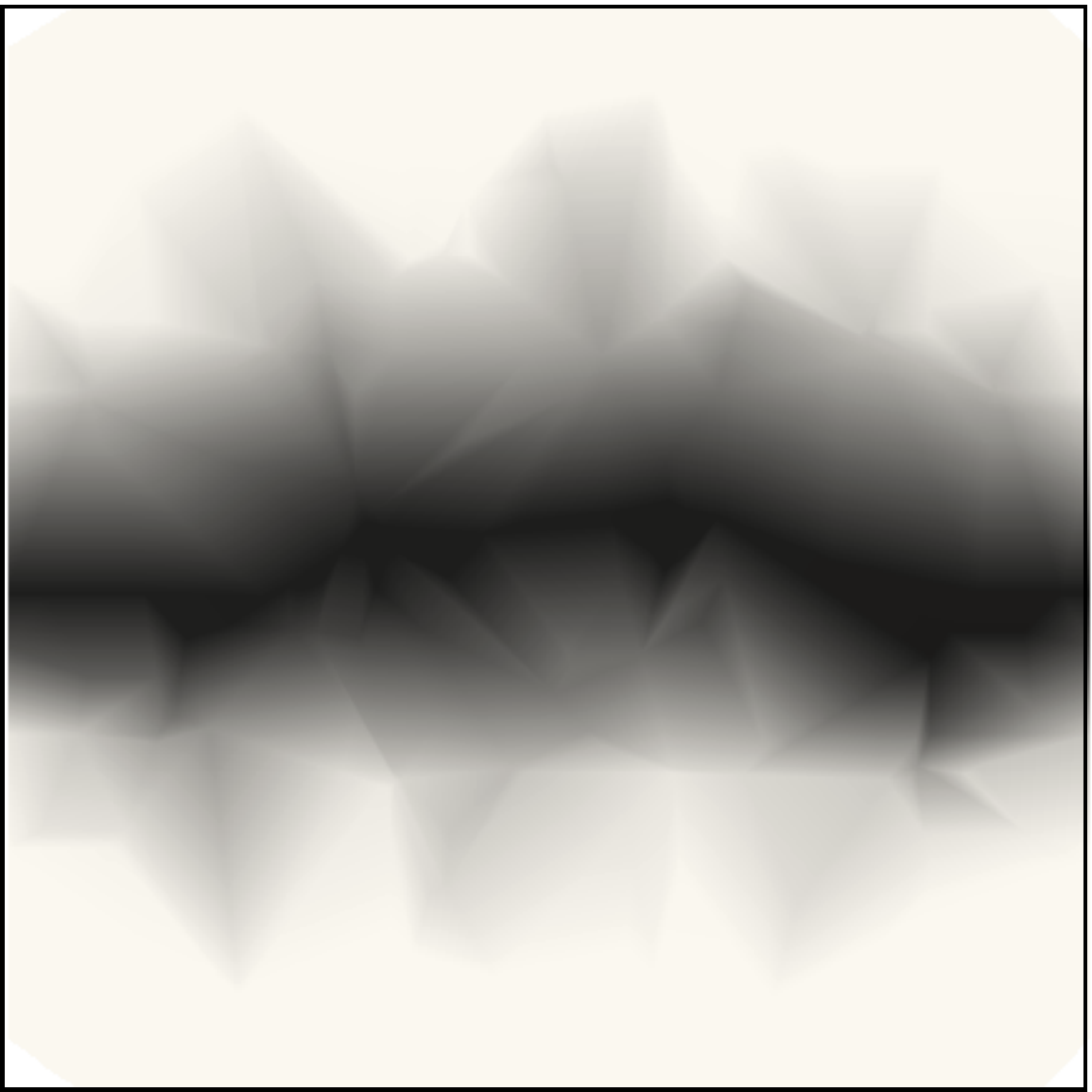}\\
     medium\\
     \includegraphics[width=11cm]{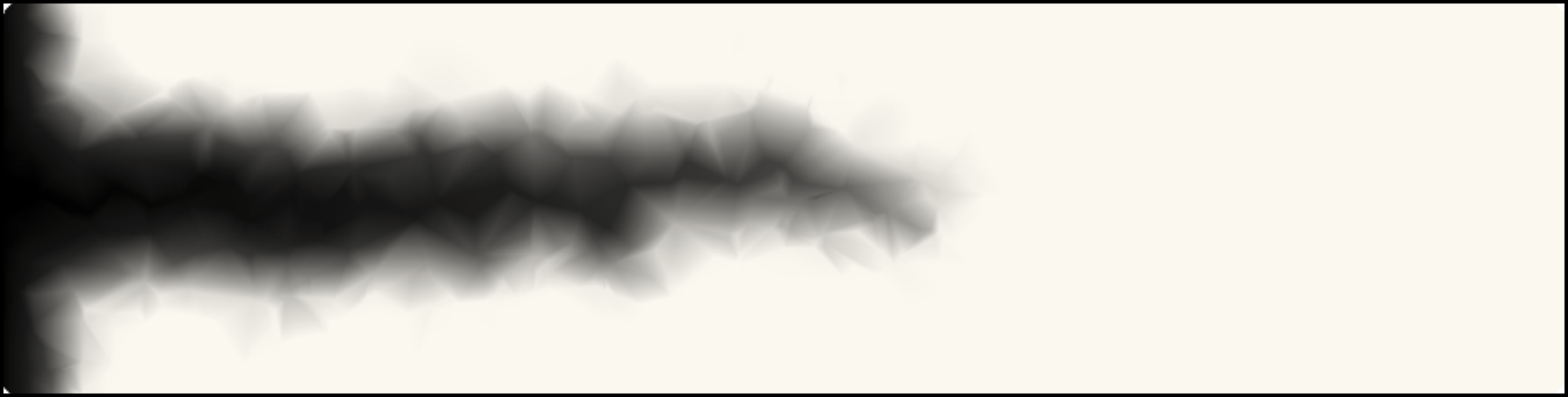} & \includegraphics[width=3cm]{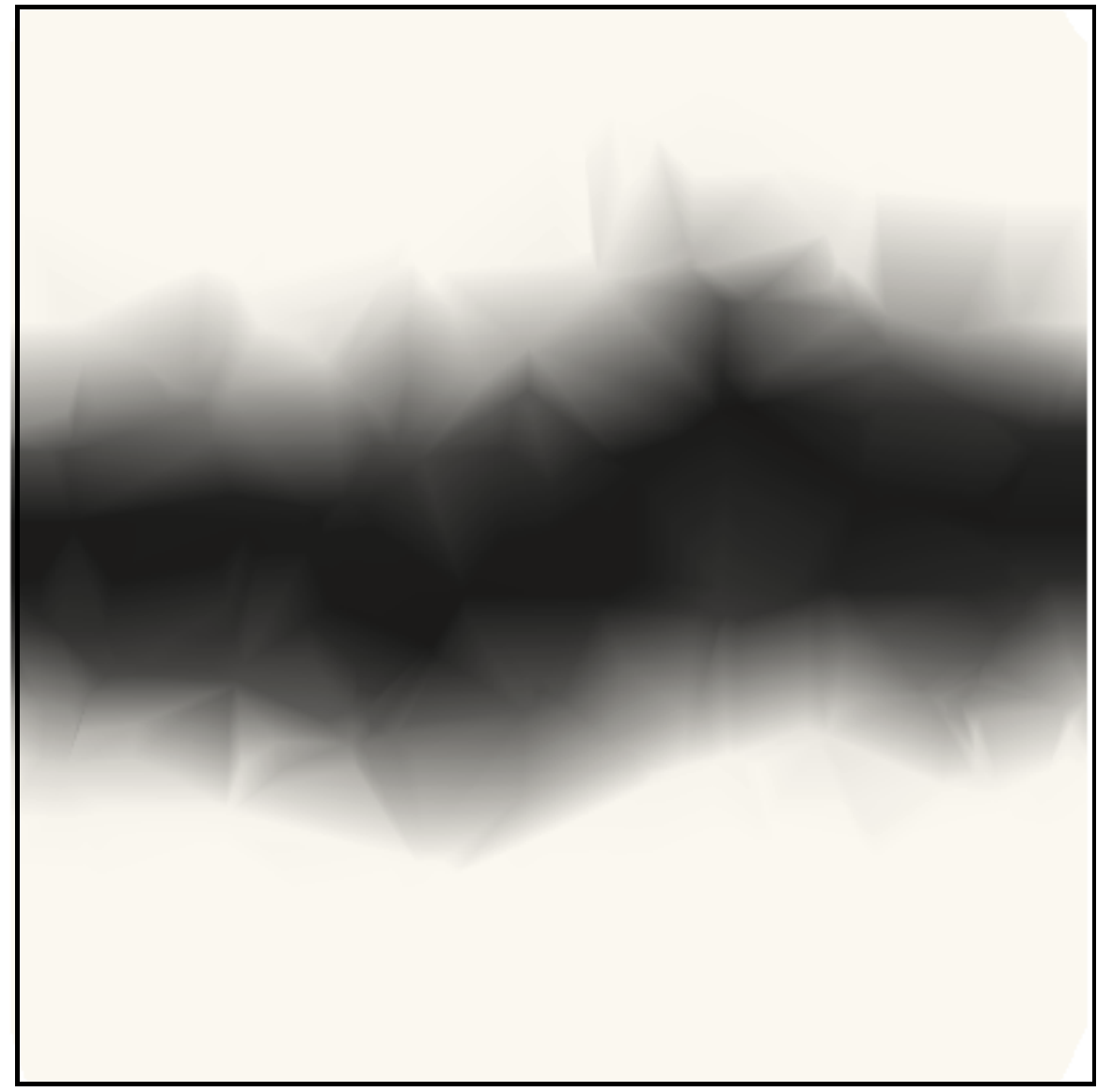}\\
     fine\\
     (\textbf{a}) &(\textbf{b}) 
\end{tabular}
\caption{Contour plots of capillary suction $P_{\rm c}$ at 3.33~h for the (\textbf{a}) $x$--$z$ plane at $y=0.125$~m and (\textbf{b})~$y$--$z$~plane at $x=0.3$~m.}
\label{fig:Contour}
\end{figure}

Furthermore, contour plots of the capillary suction $P_{\rm c}$ are shown for the three networks for the $x$--$z$ plane (at $y=0.125$~m) and for the $y$--$z$ plane (at $x = 0.3$~m) in Figure~\ref{fig:Contour}. Darker regions correspond to lower values of capillary suction, which indicate higher amounts of intruded water. Slight broadening of the intrusion zone, lateral to the crack direction, is expected for the coarser network design. Otherwise, the network model simulates the transport field element size independently. 

Whereas this example involves mode I fracture, the scalar damage model presented in Section~
\ref{sec:structuralMaterial} allows for damage development under more general loading patterns. Modification~of conductivity to account for fracture, according to Equation (\ref{eq:conSplit}) and Equation (\ref{eq:crackDiff}), is appropriate when the crack is open \mbox{(i.e., when $w_{n} > 0$)}. In this sense, the proposed model should be applicable to cases of mixed-mode loading within the tension-shear regime. Residual influences after crack closure, or possible modification of conductivity due to damage in the compression-shear regime, require additional study.    

\section{Conclusions}
A new three-dimensional network approach for modelling the effect of fracture on mass transport has been proposed.
The Delaunay tessellation of an unstructured set of points defines the structural network, which represents material elasticity and fracture. The edges of the corresponding Voronoi diagram define the network of transport elements, which simulate mass transport. A distinctive feature of the dual network approach is the alignment of transport elements with potential pathways for crack propagation. Several benchmark comparisons have been presented involving non-stationary transport, fracture, and their coupling. The following conclusions and remarks can be made.
\begin{itemize}
\item The network of structural elements, defined by the Delaunay edges, provides element geometry and size independent load-displacement curves, as demonstrated through cohesive fracture simulations of double cantilever beams. The traction free condition is approached without stress locking. Local deviations of the fracture path due to random network generation has very little influence on the load-displacement curves.   
\item The network of transport elements, defined by the Voronoi edges, provides results for non-stationary transport which are in very good agreement with analytical solutions, and are independent of  element geometry and size. The proposed discretisation scheme for the transport network facilitates the enforcement of boundary conditions. Local to a domain boundary, transport elements have one node on the boundary and are directed perpendicular to the~boundary.
\item The proposed method for coupling the effect of crack opening, determined by the structural network, with transport properties of the transport network yields objective results with respect to element geometry and size. This dual network approach facilitates the simulation of transport along crack paths and from crack faces into the bulk material. 
\end{itemize}

The proposed coupling is limited to the effect of fracture on transport. A two-way coupling of field quantities (i.e., including the dependence of structural behaviour on the transport field~\citep{GraFahGal15}) is a natural extension of this work.

\vspace{6pt}

\section*{Acknowledgments}
Peter Grassl acknowledges funding received from the UK Engineering and Physical Sciences Research Council (EPSRC) under grant EP/I036427/1 and funding from Radioactive Waste Management Limited (RWM) (http://www.nda.gov.uk/rwm), a wholly-owned subsidiary of the Nuclear Decommissioning Authority. RWM is committed to the open publication of such work in peer reviewed literature, and welcomes e-feedback to rwmdfeedback@nda.gov.uk.

\bibliographystyle{natbib}

\begin{thebibliography}{-------}
\providecommand{\natexlab}[1]{#1}

\bibitem[Roels \em{et~al.}(2006)Roels, Moonen, De~Proft, and
  Carmeliet]{RoeMooDeProCar06}
Roels, S.; Moonen, P.; de~Proft, K.; Carmeliet, J.
\newblock A coupled discrete-continuum approach to simulate moisture effects on
  damage processes in porous materials.
\newblock {\em Comput. Methods  Appl. Mech. Eng.} {\bf
  2006}, {\em 195},~7139--7153.

\bibitem[Segura and Carol(2008{\natexlab{a}})]{SegCar08}
Segura, J.M.; Carol, I.~Coupled {HM} analysis using zero-thickness interface elements with
  double nodes---{Part I: T}heoretical model.
\newblock {\em Int. J. Numer. Anal. Methods Geomech.} {\bf 2008}, {\em 32},~2083--2101.

\bibitem[Segura and Carol(2008{\natexlab{b}})]{SegCar08a}
Segura, J.M.; Carol, I.~Coupled {HM} analysis using zero-thickness interface elements with
  double nodes---{Part II: V}erification and application.
\newblock {\em Int. J. Numer. Anal. Methods   Geomech.} {\bf 2008}, {\em 32},~2103--2123.

\bibitem[Carrier and Granet(2012)]{CarGra12}
Carrier, B.; Granet, S.
\newblock Numerical modeling of hydraulic fracture problem in permeable medium
  using cohesive zone model.
\newblock {\em Eng. Fract. Mech.} {\bf 2012}, {\em 79},~312--328.

\bibitem[Yao \em{et~al.}(2015)Yao, Li, and Keer]{YaoLiKee15}
Yao, Y.; Li, L.; Keer, L.
\newblock Pore pressure cohesive zone modeling of hydraulic fracture in
  quasi-brittle rocks.
\newblock {\em Mech. Mater.} {\bf 2015}, {\em 83},~17--29.

\bibitem[Sadouki and van Mier(1997)]{SadMie97}
Sadouki, H.; van Mier, J.G.M.
\newblock Simulation of hygral crack growth in concrete repair system.
\newblock {\em Mater. Struct.} {\bf 1997}, {\em 203},~518--526.

\bibitem[Chatzigeorgiou \em{et~al.}(2005)Chatzigeorgiou, Picandet, Khelidj, and
  Pijaudier-Cabot]{ChaPicKhePij05}
Chatzigeorgiou, G.; Picandet, V.; Khelidj, A.; Pijaudier-Cabot, G.
\newblock {Coupling between progressive damage and permeability of concrete:
  analysis with a discrete model}.
\newblock {\em Int. J. Numer. Anal. Methods Geomech.} {\bf 2005}, {\em 29},~1005--1018.

\bibitem[Nakamura \em{et~al.}(2006)Nakamura, Srisoros, Yashiro, and
  Kunieda]{NakSriYas06}
Nakamura, H.; Srisoros, W.; Yashiro, R.; Kunieda, M.
\newblock Time-dependent structural analysis considering mass transfer to
  evaluate deterioration process of {RC} structures.
\newblock {\em J. Adv. Concr. Technol.} {\bf 2006}, {\em
  4},~147--158.

\bibitem[Wang \em{et~al.}(2008)Wang, Soda, and Ueda]{WanSodUed08}
Wang, L.; Soda, M.; Ueda, T.
\newblock Simulation of chloride diffusivity for cracked concrete based on
  {RBSM} and truss network model.
\newblock {\em J. Adv. Concr. Technol.} {\bf 2008}, {\em
  6},~143--155.

\bibitem[Grassl(2009)]{Gra09}
Grassl, P.
\newblock A lattice approach to model flow in cracked concrete.
\newblock {\em Cem. Concr. Compos.} {\bf 2009}, {\em 31},~454--460.

\bibitem[\v{S}avija \em{et~al.}(2013)\v{S}avija, Pacheco, and
  Schlangen]{SavPacSch13}
\v{S}avija, B.; Pacheco, J.; Schlangen, E.
\newblock Lattice modeling of chloride diffusion in sound and cracked concrete.
\newblock {\em Cem. Concr. Compos.} {\bf 2013}, {\em 42},~30--40.

\bibitem[Asahina \em{et~al.}(2014)Asahina, Houseworth, Birkholzer, Rutqvist,
  and Bolander]{AsaHouBir14}
Asahina, D.; Houseworth, J.E.; Birkholzer, J.T.; Rutqvist, J.; Bolander, J.E.
\newblock Hydro-mechanical model for wetting/drying and fracture development in
  geomaterials.
\newblock {\em Comput. Geosci.} {\bf 2014}, {\em 65},~13--23.

\bibitem[Grassl \em{et~al.}(2015)Grassl, Fahy, Gallipoli, and
  Wheeler]{GraFahGal15}
Grassl, P.; Fahy, C.; Gallipoli, D.; Wheeler, S.J.
\newblock On a {2D} hydro-mechanical lattice approach for modelling hydraulic
  fracture.
\newblock {\em J. Mech. Phys. Solids} {\bf 2015}, {\em
  75},~104--118.

\bibitem[Marina \em{et~al.}(2015)Marina, Derek, Mohamed, Yong, and
  Imo-Imo]{MarDerMoh15}
Marina, S.; Derek, I.; Mohamed, P.; Yong, S.; Imo-Imo, E.K.
\newblock Simulation of the hydraulic fracturing process of fractured rocks by
  the discrete element method.
\newblock {\em Environ. Earth Sci.} {\bf 2015}, {\em 73},~8451--8469.

\bibitem[Damjanac \em{et~al.}(2015)Damjanac, Detournay, and
  Cundall]{DamDetCun15}
Damjanac, B.; Detournay, C.; Cundall, P.A.
\newblock Application of particle and lattice codes to simulation of hydraulic
  fracturing.
\newblock {\em Comput. Part. Mech.} {\bf 2015}, pp. 1--13.

\bibitem[Bolander and Saito(1998)]{BolSai98}
Bolander, J.E.; Saito, S.
\newblock Fracture analysis using spring networks with random geometry.
\newblock {\em Eng. Fract. Mech.} {\bf 1998}, {\em 61},~569--591.

\bibitem[Bolander and Sukumar(2005)]{BolSuk05}
Bolander, J.E.; Sukumar, N.  Irregular lattice model for quasistatic crack propagation.
 {\em Phys. Rev. B} {\bf 2005}, {\em 71},~094106.

\bibitem[Grassl and Jir\'{a}sek(2010)]{GraJir10}
Grassl, P.; Jir\'{a}sek, M.
\newblock Meso-scale approach to modelling the fracture process zone of
  concrete subjected to uniaxial tension.
\newblock {\em Int. J. Solids Struct.} {\bf 2010}, {\em
  47},~957--968.

\bibitem[Grassl and Davies(2011)]{GraDav11}
Grassl, P.; Davies, T.
\newblock Lattice modelling of corrosion induced cracking and bond in
  reinforced concrete.
\newblock {\em Cem.~Concr. Compos.} {\bf 2011}, {\em 33},~918--924.

\bibitem[Bolander and Berton(2004)]{BolBer04}
Bolander, J.E.; Berton, S.
\newblock Simulation of shrinkage induced cracking in cement composite
  overlays.
\newblock {\em Cem.~Concr. Compos.} {\bf 2004}, {\em 26},~861--871.

\bibitem[Saka(2012)]{Saka12}
Saka, T.
\newblock Simulation of Reinforced Concrete Durability: Dual-Lattice Models of
Crack-Assisted Mass Transport.
\newblock Ph.D. Thesis, University of California, Davis, CA, USA,  2012.

\bibitem[Okabe \em{et~al.}(2000)Okabe, Boots, Sugihara, and Chiu]{OkaBooSug00}
Okabe, A.; Boots, B.; Sugihara, K.; Chiu, S.N.
\newblock {\em Spatial Tessellations: Concepts and Applications of Voronoi
  Diagrams}; Wiley: New York, NY, USA, 2000.

\bibitem[Yip \em{et~al.}(2005)Yip, Mohle, and Bolander]{YipMohBol05}
Yip, M.; Mohle, J.; Bolander, J.E.
\newblock {Automated modeling of three-dimensional structural components using
  irregular lattices}.
\newblock {\em Comput. Aided Civ.  Infrastruct. Eng.} {\bf 2005},
  {\em 20},~393--407.

\bibitem[Strang(1986)]{Strang86}
Strang, G.
\newblock {\em Introduction to Applied Mathematics}; Wellesley-Cambridge Press:
  Wellesley, MA, USA,  1986.

\bibitem[Kawai(1978)]{Kaw78}
Kawai, T. New discrete models and their application to seismic response
  analysis of structures. \newblock{\em Nucl. Eng. Des.} {\bf 1978}, {\em 48},~207--229.

\bibitem[Berton and Bolander(2006)]{BerBol06}
Berton, S.; Bolander, J.E.
\newblock Crack band model of fracture in irregular lattices.
\newblock {\em Comput. Methods Appl. Mech.~Eng.} {\bf
  2006}, {\em 195},~7172--7181.

\bibitem[Mazars and Pijaudier-Cabot(1989)]{MazPij89}
Mazars, J.; Pijaudier-Cabot, G.
\newblock Continuum damage theory---Application to concrete.
\newblock {\em J. Eng. Mech.} {\bf 1989}, {\em 115},~345--365.

\bibitem[Lemaitre and Chaboche(1990)]{Lemaitre90}
Lemaitre, J.; Chaboche, J.L.
\newblock {\em Mechanics of solid materials}; Cambridge University Press:
  Cambridge, UK,  1990.

\bibitem[Maekawa \em{et~al.}(2008)Maekawa, Ishida, and Kishi]{MaeIshKis08}
Maekawa, K.; Ishida, T.; Kishi, T.
\newblock {\em Multi-Scale Modeling of Structural Concrete}; CRC Press: Boca Raton, FL, USA, 2008.

\bibitem[Lewis \em{et~al.}(1996)Lewis, Morgan, Thomas, and
  Seetharamu]{LewMorTho96}
Lewis, R.W.; Morgan, K.; Thomas, H.R.; Seetharamu, K.
\newblock {\em The Finite Element Method in Heat Transfer Analysis}; John Wiley
  \& Sons: Hoboken, NJ, USA, 1996.

\bibitem[{van Genuchten}(1980)]{Gen80}
{van Genuchten}, M.T.
\newblock A closed-form equation for predicting the hydraulic conductivity of
  unsaturated soils.
\newblock {\em Soil Sci. Soc. Am.} {\bf 1980}, {\em 44},~892--898.

\bibitem[Baroghel-Bouny \em{et~al.}(1999)Baroghel-Bouny, Mainguy, Lassabatere,
  and Coussy]{BarMaiLas99}
Baroghel-Bouny, V.; Mainguy, M.; Lassabatere, T.; Coussy, O.
\newblock Characterization and identification of equilibrium and transfer
  moisture properties for ordinary and hig{h-p}erformance cementitious
  materials.
\newblock {\em Cem. Concr. Res.} {\bf 1999}, {\em 29},~1225--1238.

\bibitem[Akhavan \em{et~al.}(2012)Akhavan, Shafaatian, and
  Rajabipour]{AkhShaRaj12}
Akhavan, A.; Shafaatian, S.M.H.; Rajabipour, F.
\newblock Quantifying the effects of crack width, tortuosity, and roughness on
  water permeability of cracked mortars.
\newblock {\em Cem. Concr. Res.} {\bf 2012}, {\em 42},~313--320.

\bibitem[Witherspoon \em{et~al.}(1980)Witherspoon, Wang, Iawai, and
  Gale]{WitWanIwaGal80}
Witherspoon, P.A.; Wang, J.S.Y.; Iawai, K.; Gale, J.E.
\newblock Validity of cubic law for fluid flow in a deformable rock fracture.
\newblock {\em Water Resour. Res.} {\bf 1980}, {\em 16},~1016--1024.

\bibitem[Asahina \em{et~al.}(2011)Asahina, Landis, and Bolander]{AsaLanBol11}
Asahina, D.; Landis, E.N.; Bolander, J.E.
\newblock Modeling of phase interfaces during pre-critical crack growth in
  concrete.
\newblock {\em Cem. Concr. Compos.} {\bf 2011}, {\em 33},~966--977.

\bibitem[Patz\'ak(2012)]{Pat12}
Patz\'ak, B.
\newblock {OOFEM}---An object-oriented simulation tool for advanced modeling
  of materials and structures.
\newblock {\em Acta Polytech.} {\bf 2012}, {\em 52},~59--66.

\end{thebibliography}

\renewcommand\bibname{References}

\end{document}